\documentclass[conference]{IEEEtran}
\IEEEoverridecommandlockouts
\usepackage{booktabs}
\usepackage{graphicx}
\usepackage{cite}
\usepackage{amsmath,amssymb,amsfonts}
\usepackage{algorithmic}
\usepackage{graphicx}
\usepackage{textcomp}
\usepackage{xcolor}
\usepackage{url}
\usepackage{footnote}
\usepackage{todonotes}
\usepackage{amsmath,amssymb,amsfonts}
\usepackage{scalerel}
\usepackage{tikz}
\usetikzlibrary{svg.path}
\def\BibTeX{{\rm B\kern-.05em{\sc i\kern-.025em b}\kern-.08em
    T\kern-.1667em\lower.7ex\hbox{E}\kern-.125emX}}

\input{style/listings.cls}

\definecolor{orcidlogocol}{HTML}{A6CE39}
\tikzset{
  orcidlogo/.pic={
    \fill[orcidlogocol] svg{M256,128c0,70.7-57.3,128-128,128C57.3,256,0,198.7,0,128C0,57.3,57.3,0,128,0C198.7,0,256,57.3,256,128z};
    \fill[white] svg{M86.3,186.2H70.9V79.1h15.4v48.4V186.2z}
                 svg{M108.9,79.1h41.6c39.6,0,57,28.3,57,53.6c0,27.5-21.5,53.6-56.8,53.6h-41.8V79.1z M124.3,172.4h24.5c34.9,0,42.9-26.5,42.9-39.7c0-21.5-13.7-39.7-43.7-39.7h-23.7V172.4z}
                 svg{M88.7,56.8c0,5.5-4.5,10.1-10.1,10.1c-5.6,0-10.1-4.6-10.1-10.1c0-5.6,4.5-10.1,10.1-10.1C84.2,46.7,88.7,51.3,88.7,56.8z};
  }
}

\newcommand\orcidicon[1]{\href{https://orcid.org/#1}{\mbox{\scalerel*{
\begin{tikzpicture}[yscale=-1,transform shape]
\pic{orcidlogo};
\end{tikzpicture}
}{|}}}}

\usepackage{listings,color}
\usepackage{hyperref} 

\begin{document}

\title{A Tutorial on the Interoperability of Self-sovereign Identities \\
\thanks{This paper has been realized via funding from the IDunion project by German Ministry of Economic Affairs and Climate Action (BMWK), grant number 01MN21002K.The information and views set out in this publication are those of the authors and do not necessarily reflect the official opinion of the BMWK. Responsibility for the information and views
expressed here lies entirely with the authors.}
}

\author{\IEEEauthorblockN{Hakan Yildiz \orcidicon{0000-0002-2044-0826}}
\IEEEauthorblockA{\textit{TU Berlin, Service-centric Networking}\\
Berlin, Germany \\
hakan.yildiz@tu-berlin.de}
\and
\IEEEauthorblockN{Axel Küpper \orcidicon{0000-0002-4356-5613}}
\IEEEauthorblockA{\textit{TU Berlin, Service-centric Networking}\\
Berlin, Germany \\
axel.kuepper@tu-berlin.de}
\and
\IEEEauthorblockN{Dirk Thatmann \orcidicon{0000-0002-6646-281X}}
\IEEEauthorblockA{\textit{Telekom Innovation Laboratories (T-Labs)}\\
Berlin, Germany \\
dirk.thatmann@telekom.de}
\and
\IEEEauthorblockN{\hspace{3.3cm} Sebastian Göndör \orcidicon{0000-0003-0402-4379}}
\IEEEauthorblockA{\textit{\hspace{3.3cm}TU Berlin, Service-centric Networking}\\
\hspace{3.3cm}Berlin, Germany \\
\hspace{3.3cm}sebastian.goendoer@tu-berlin.de}
\and
\IEEEauthorblockN{\hspace{-3.3cm}Patrick Herbke \orcidicon{0000-0001-9649-2975}}
\IEEEauthorblockA{\textit{\hspace{-3.3cm}TU Berlin, Service-centric Networking}\\
\hspace{-3.3cm}Berlin, Germany \\
\hspace{-3.3cm}p.herbke@tu-berlin.de}
}


\maketitle

\begin{abstract}
Self-sovereign identity is the latest digital identity paradigm that allows users, organizations, and things to manage identity in a decentralized fashion without any central authority controlling the process of issuing identities and verifying assertions. Following this paradigm, implementations have emerged in recent years, with some having different underlying technologies. These technological differences often create interoperability problems between software that interact with each other from different implementations. Although a common problem, there is no common understanding of self-sovereign identity interoperability. In the context of this tutorial, we create a definition of interoperability of self-sovereign identities to enable a common understanding. Moreover, due to the decentralized nature, interoperability of self-sovereign identities depends on multiple components, such as ones responsible for establishing trust or enabling secure communication between entities without centralized authorities. To understand those components and their dependencies, we also present a reference model that maps the required components and considerations that build up a self-sovereign identity implementation. The reference model helps address the question of how to achieve interoperability between different implementations. 
\end{abstract}


\section{Introduction}
Digital identities are sets of claims made by the identity subject about itself or another entity \cite{identity}, where a claim is a simple statement about the identity subject or another entity. For example, a claim can be a simple username or a set of attributes such as eye color, height, and age that describe the subject.
Identity subjects can have multiple digital identities for different purposes.
Generally speaking, possessing multiple digital identities is encouraged as they can be asserted for different purposes without disclosing irrelevant or personal information (claims).
For instance, one digital identity might be asserted for accessing online banking and another for an online social network: The bank does not need to know the identity related to the online social network or vice versa. 

The primary purpose of a digital identity is to enable authentication and authorization of the identity subject for accessing goods and services. Cameron \cite{identity} also suggests that a system processing digital identities must reliably extract the identifying information of the identity subject to another subject such as a service provider (SP) while detecting deception attempts such as replay attacks. For this purpose, during the early stages of the Internet, SPs had their identity access management systems (IDMS) and identity providers (IDP). This paradigm is called siloed identities, which are still popular and in use. For instance, online banking services generally allow access by identities created and stored in the IDPs belonging to the same bank. While siloed identities offer benefits such as one identity for one service and limited impact in case of identity theft, users end up with countless service-specific identities.

Some IDPs became so large that they started offering authentication for other SPs outside their domain. This type of one IDP to many SP relation created the centralized identities paradigm. It is also known as single sign-on (SSO). Examples of SSO are Google and Facebook login, which can be used for services other than Gmail and Facebook. SSO is proven to be an improvement in convenience and user experience. However, it also came with some disadvantages. First, these large IDPs became honeypots for identity theft. Since one identity is used for multiple services, the loss of this one identity is proportionally more impactful than siloed identities. Second, if the IDP denies authentication to the identity subject, he cannot access any of the services from the SPs relying on the authentication from the denying IDP. 

Later on, centralized IDPs and SPs started creating a circle of trust by SPs accepting identities from the IDPs in the circle, creating many IDPs to many SPs relation, which is known as the federated identity paradigm. Federated identities solve some issues related to centralized identities, such as dependency on a single IDP. However, the federation comes with two issues. First, participants of the federation must agree on a set of policies, creating governance overhead for the federation. Second, participating IDPs and SPs must be interoperable by design. Therefore, they must align on the standards for the assertion of identity. Commonly known standards are SAML\cite{lockhart2008security}, OAuth\cite{OAUTH_RFC}, and OpenID Connect\cite{OIDC_SPEC}. Centralized and federated identities are also defined as user-centric identities when the identity subject can consent to the shared data with the SP. 

Self-sovereign identity (SSI) is the latest digital identity paradigm, enabling identity subjects to control and own their digital identities. In contrast to other identity paradigms, self-sovereign identities exist outside of IDP within the domain of the identity subject. Due to this property, the subject is often referred to as the identity holder. With SSI, the role of an IDP is reduced to an issuer as it no longer stores the digital identity or asserts the authentication result to an SP. Instead, a self-sovereign identity digitally signed by an issuer can be validated with key material belonging to the issuer that is stored in highly available and tamper-proof registries. Moreover, in many cases, an issuer is not required as the identity subject can create a new self-sovereign identity to access a service from an SP that does not require official identification.

\begin{figure}[!ht]
	\centering
	\includegraphics[width=85mm]{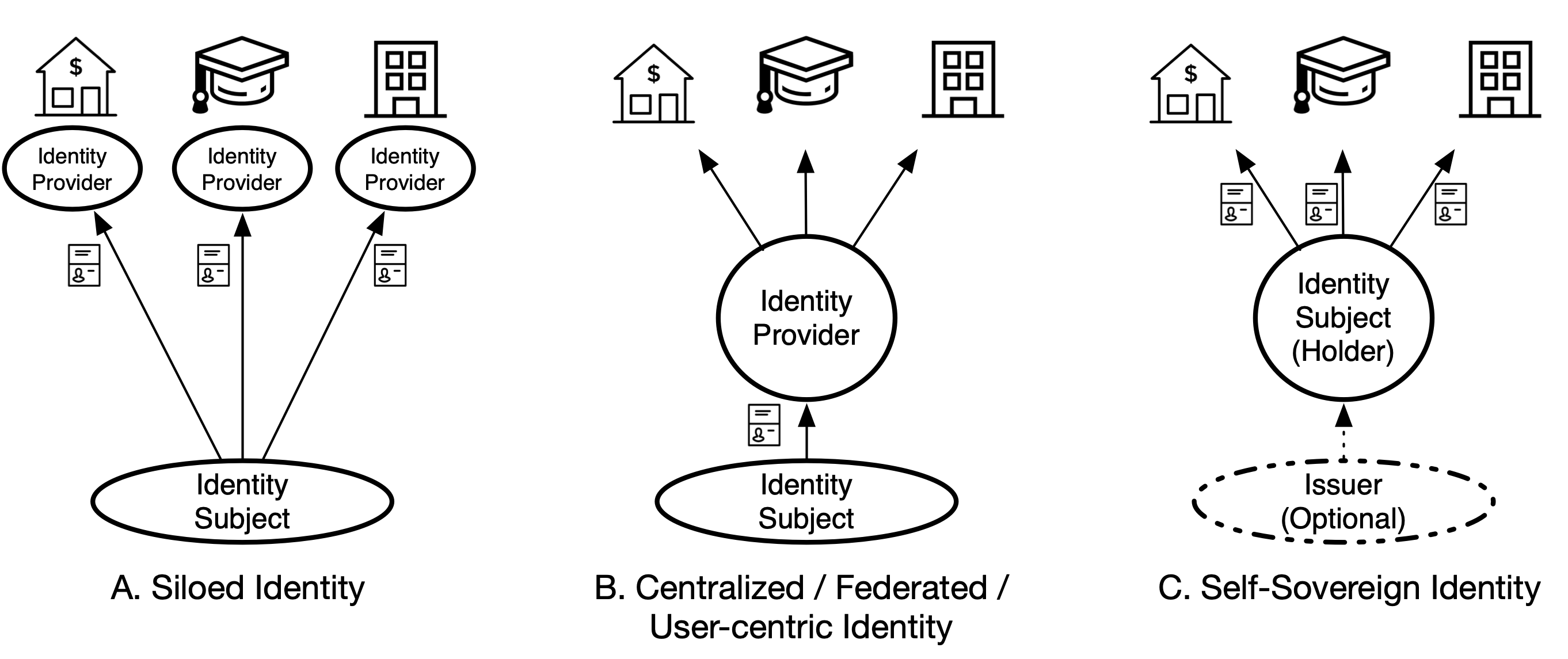}\\
	\caption{Digital Identity Paradigms adapted from \cite{Yild2109:Connecting}}
	\label{fig:identity_paradigms}
\end{figure}

Figure \ref{fig:identity_paradigms} visualizes the differences between the digital identity paradigms.

Like other identity paradigms mentioned above, SSI comes with its own challenges. There are multiple implementations that follow the SSI approach while using different technologies. Implementations with different technologies are incompatible and, therefore, not interoperable with each other. Similar to federated identities, the acceptance can be achieved if the issuers, holders, and SPs can align on the standards of the assertion of identities. However, the magnitude of differences in SSI implementations is significantly higher than in federated identities, which focus on standardization of assertion. Since SSI is a decentralized approach and the identities can exist outside of IDPs, SSI implementations require not only standardization of assertion but also the creation of trust and communication. Therefore, achieving interoperability is also more complex than the federated approach. 

Although there is a need for interoperability between different SSI implementations, there is no common understanding between organizations to set interoperability goals. Moreover, due to its decentralized approach and the necessity of decentralized infrastructures, and therefore, its significant differences from the other digital identity paradigms, the understanding of how to achieve the set interoperability goals is equally missing. The prior requires a definition, while the latter needs a systematic approach to understand the differences between implementations.

In this tutorial, we present a definition of SSI interoperability to create a shared understanding of what interoperability in the context of SSI means. The definition is based on a model accepted by standards developing organizations, which can enable entities with different implementations to set interoperability goals. Moreover, we present a reference model comprising multiple components categorized in layers. The reference model can help understand the differences between SSI implementations, which ultimately helps address the question of how to reach the interoperability goals.

The rest of this tutorial is structured as follows: In Section \ref{SSIbasics}, we give a brief overview of identity trust and its depiction in different identity paradigms. In Section \ref{problem} we define the problem regarding SSI interoperability. Then we discuss the definition of SSI interoperability in Section \ref{taxonomy} and the reference model in Section \ref{layers} to understand the differences between implementations. Based on the reference model, we discuss the interoperability considerations in Section \ref{considerations}. Finally, we conclude this tutorial in Section \ref{conclude}.

\section{Background} \label{SSIbasics}

\begin{figure}[!ht]
	\centering
	\includegraphics[width=85mm]{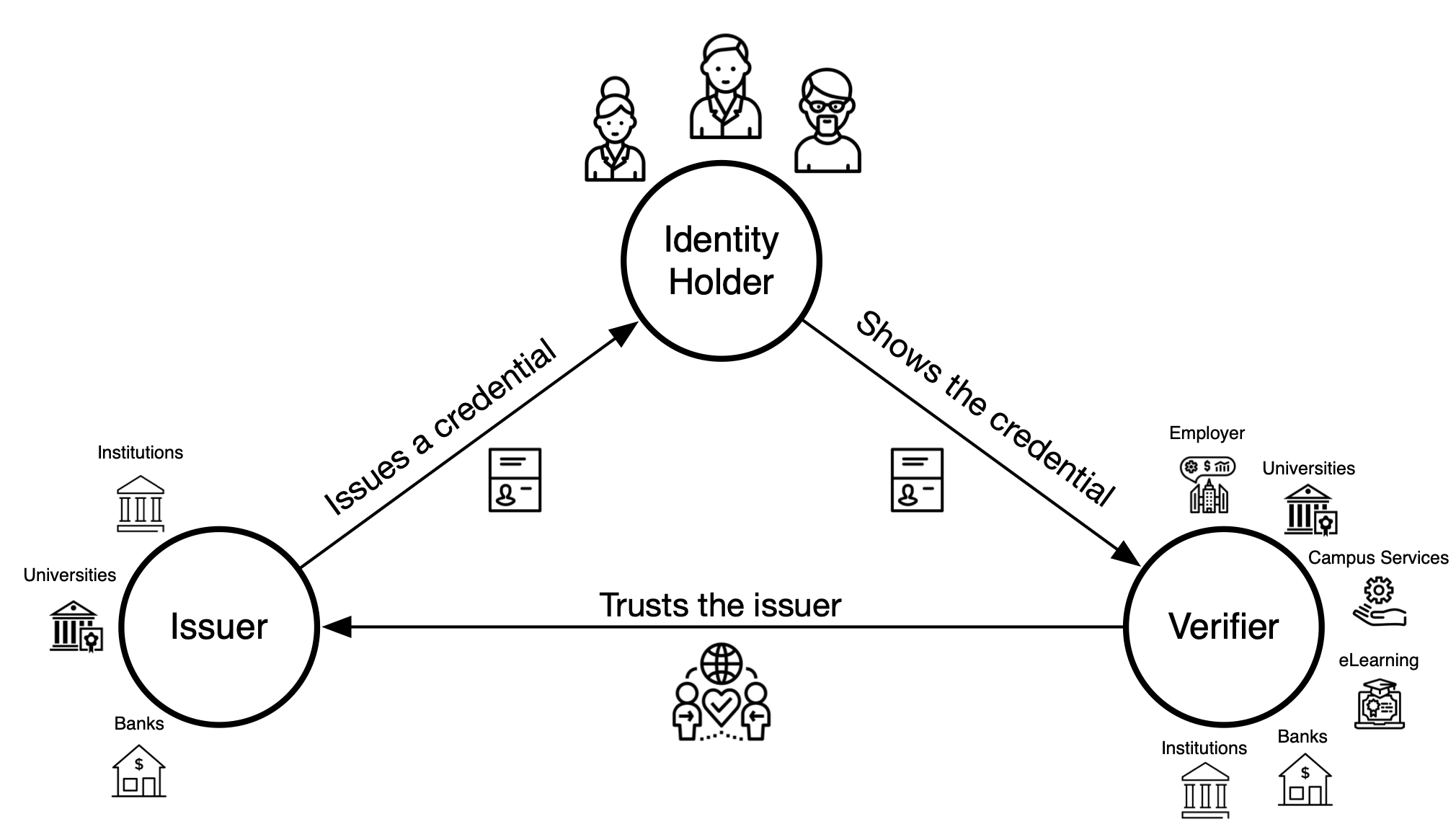}\\
	\caption{Identity Trust Triangle}
	\label{trust_triangle}
\end{figure}

Like any digital or physical identity paradigm, SSI is based on the identity trust triangle that describes the interactions between three actors: an issuer, an identity holder, and a verifier. The issuer is responsible for creating credentials that contain one or more claims about the identity holder, which are verifiable. In siloed, centralized, and federated identities, the issuer instance is also known as the IDP. Moreover, the identity holder is the owner of the credential. When identity holders want to access goods and services, they present a credential to prove they are authorized. The final actor is the verifier, responsible for ensuring that the presented credentials are valid and the holder is authorized for the transaction. Generally, verifiers are SPs. The triangle is completed with the verifier having either a direct or indirect trust in the issuer. Figure \ref{trust_triangle} visualizes the identity trust triangle.

\subsection{Trust Triangle in Physical Identities}
For example, Mike is a student at TU Berlin (issuer). He acquired his student ID from his university as he started to study there (identity holder). Mike wants to go to a bar that is having a student night: it offers a 50\% discount on beverages to students that night. Mike wants to order a beverage (accessing goods and services). The bartender asks for an identification (SP and verifier) before giving him a discounted beverage. Mike presents his student ID to the bartender. The bartender makes a visual inspection. Based on the visual authenticity and his trust in the issuing party, he gives Mike the ordered beverage with a 50\% discount.

\subsection{Trust Triangle in Digital Identities}
The same principle can be applied to digital identities. As Mike started his studies, his campus management system also provisioned him with a service account that he could access with a username and password. Mike is writing a paper and searching for publications in IEEE. Mike finds a paper related to his paper and wants to read the full version. To do so, IEEE requires either a payment or a proof of authorization that he can access the full version of the paper without paying additionally. To that end, IEEE relies on a federated IDMS and accepts credentials from multiple universities. Mike chooses his university on the login page, and he is redirected to the sign-in page of his university. He provides his username and password, which are then validated by the campus IDP and the result is sent back. IEEE trusts the IDP in the federated system and grants Mike access to the paper. 

\subsection{Trust Triangle in Self-sovereign Identities}
SSI is the latest digital identity paradigm that enables the trust relationship between actors according to the trust triangle. This time, similar to the physical identities, an issuer creates a self-sovereign identity that the identity holder can store at his medium of choice. Self-sovereign identities are digitally signed with a private key of a public/private key pair belonging to the issuer. If the content is tampered with, the signature value does not fit the content anymore. Therefore, self-sovereign identities are tamper-proof, and their origins are cryptographically verifiable.

An identity holder can present a derivation of a self-sovereign identity that belongs to him. He can do so by creating a proof of ownership of the identity and presenting it as a one-time usable derivation. For example, a proof for the derivation of the self-sovereign identity can be created by signing the derivation with a private key of a public/private key-pair belonging to the identity holder. The related public key is specified in the issued self-sovereign identity. Due to the tamper-proof property of self-sovereign identities, the verifier can trust that the credential is bound to the presenting party. 

\begin{figure}[t!]
	\vspace{0.9cm}
	\centering
	\includegraphics[width=85mm]{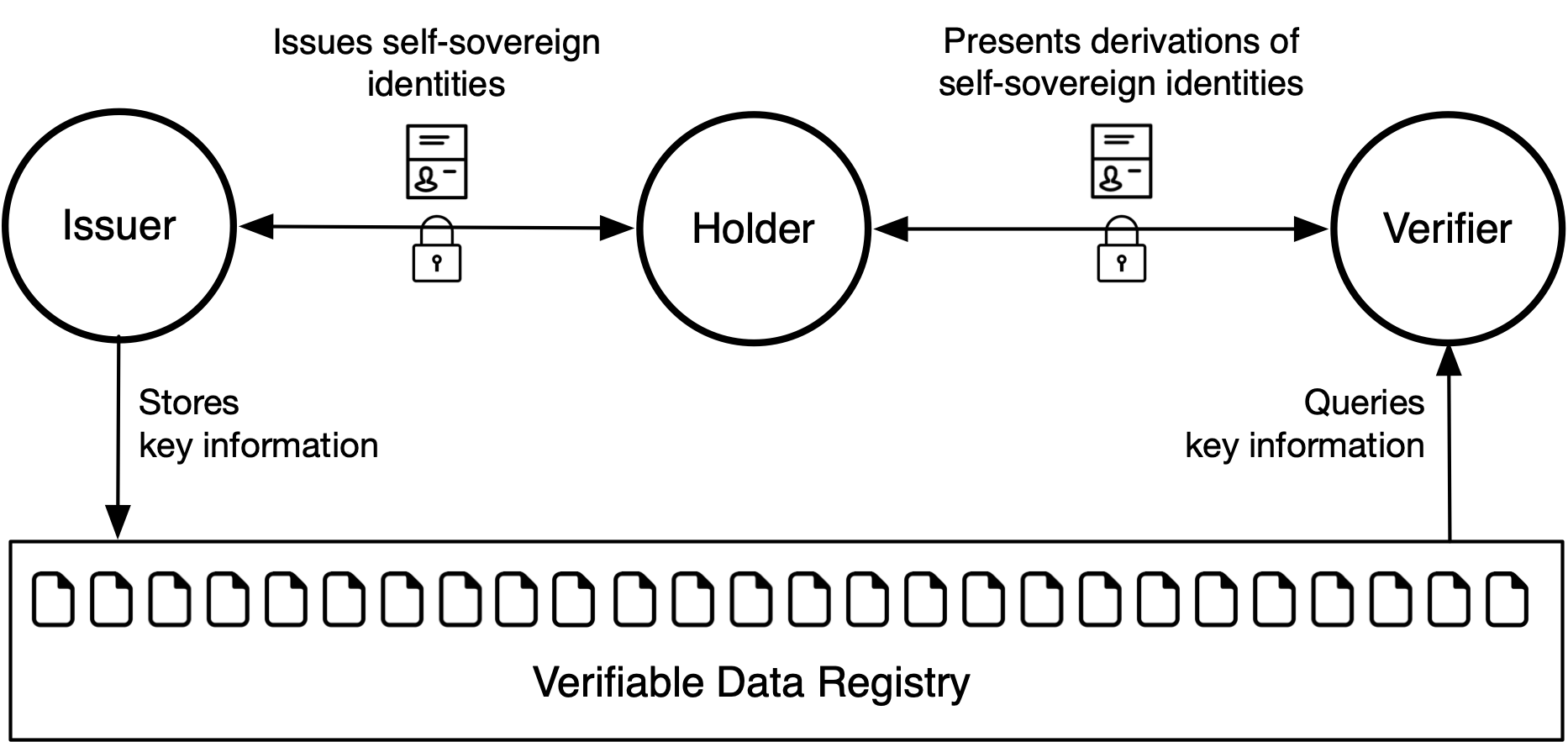}\\
	\caption{Trust relation between the SSI actors, simplified}
	\label{issuer-holder-verifier}
\end{figure}

The trust between the verifier and the issuer is indirect with the SSI paradigm. Hence, there is no direct communication between the issuer and the verifier to validate the presented derivation of self-sovereign identities. Instead, the issuer stores information about its public keys in a tamper-resistant and highly available registry called a verifiable data registry (VDR). This information is available to the verifier. During the verification process, the proof in the self-sovereign identity can be validated with the public key stored in the VDR. Since a VDR is tamper-resistant, the verifier trusts that the stored information belongs to the issuer and the public key is not tampered with, which also completes the identity trust triangle. Figure \ref{issuer-holder-verifier} depicts the actors and their interactions.

\section{Problem Statement}
\label{problem}

On an abstract level, Section \ref{SSIbasics} explains the interactions and the creation of trust between the three SSI actors, namely issuer, identity holder, and verifier. However, the implementation of self-sovereign identities can vary depending on the requirements set by the business needs. For instance, Linked Data utilization for achieving semantic disambiguity might have high importance for one organization. At the same time, another organization might prioritize privacy-preserving features or a much simpler approach that does not provide any of those characteristics. The result is a divergence in the utilized technology stacks.

For example, the OpenID Connect (OIDC) Foundation uses an extension of its current technology stack to support self-sovereign identities. It deploys JSON Web Tokens for data exchange while using a communication method that relies on the existing OIDC standards, implementations, and public-key infrastructure\cite{siopv2}. On the other hand, Digital Bazaar pioneered the utilization of Linked Data with SSI-based credentials to create semantic disambiguity\cite{json-ld}, thus enabling a shared understanding and interpretation of data while utilizing the decentralized public-key infrastructure to create trust and secure communication between actors. Similarly, the Hyperledger Indy is a technology stack that takes a similar approach as Digital Bazaar for securing trust and communication between the actors. However, it takes a different approach for the credentials that enable privacy-preserving features such as enclosing only the necessary data and predicates that allow logical expressions\cite{indyprivacy}.

Consequently, this divergence has practical implications for the interactions between actors using different technology stacks. They are not interoperable to the extent that some of the functionalities of SSI, such as creating secure communication and issuing credentials, cannot be fulfilled by every actor. As a result, isolated ecosystems rose. A group of actors who used a specific technology stack can interact, while other actors who use different stacks cannot. 

Interoperability is one of the ten principles of SSI. These principles can be perceived as the characteristics of implementations that must be included to ensure user control is at the center of SSI. Furthermore, interoperability ensures that digital identities are widely available, crossing national or corporational boundaries to achieve genuinely global identities\cite{allen2016path}. Therefore, lack of interoperability is a severe problem of self-sovereign identities that must be addressed to ensure broad adoption and acceptance.

As of the writing of this tutorial, no one solution solves the interoperability problem of self-sovereign identities. Therefore, the discovery or proposal of such a potential solution is also not within the scope of this work. Instead, we propose a systematic approach for understanding the underlying issues that hinder interoperability by defining interoperability regarding SSI, presenting a reference model to understand the SSI technology stacks, and discussing the interoperability considerations based on the reference model. In the following, we discuss these topics in detail. 

\subsection{Definition of SSI Interoperability}

Numerous implementations of self-sovereign identities are created in Europe and North America. In the United States, seven of these implementations have been funded by Silicon Valley Innovation Program (SVIP) to become interoperable under the umbrella of a supply chain use case\footnote{Multi-Platform \& Multi-Vendor Interoperability Showcase: \url{https://lists.w3.org/Archives/Public/public-credentials/2021Mar/0101.html}}. In Europe, the European Self-Sovereign Identity Framework (eSSIF) implemented a custom solution that takes parts of its technology from the OIDC and the Hyperledger stack\footnote{Architecture Diagram of EBSI V2: \url{https://ec.europa.eu/digital-building-blocks/wikis/display/EBSIDOC/Architecture}}. The goal is to apply this solution in the public sector, such as through university diplomas. In Germany, four consortia under the showcase project Secure Digital Identities are implementing self-sovereign identities with diverging technologies\footnote{Showcase programme “Secure Digital Identities”\url{https://www.digitale-technologien.de/DT/Redaktion/EN/Downloads/Publikation/sdi_showcase-programme.pdf?__blob=publicationFile&v=2}}. 

Interoperability is required in the United States via the SVIP project. It is also mentioned on a European level via eSSIF and required in Germany through the showcase project, where the implementations of four consortia must be interoperable. Interoperability is also mentioned for transatlantic use cases between Europe and North America\footnote{Transatlantic SSI Interop: \url{https://medium.com/@markus.sabadello/transatlantic-ssi-interop-52bac6be8dfe}}. Although there is a strong emphasis on interoperability between SSI implementations, a definition of SSI interoperability has not been defined or agreed upon. Therefore, it is crucial to agree on the meaning of SSI interoperability to achieve it ultimately.

\subsection{SSI Reference Model}

An SSI technology stack is a set of choice of individual components that enable all of the required functionalities of self-sovereign identities. The SSI technology stack comprises dozens of components, and multiple implementations exist for each. This modularity emphasizes the importance of common standards for interoperability. The differences between technology stacks and the options within the components must be described to comprehend the interoperability problem. Therefore, in addition to the definition of SSI interoperability, we need an abstraction such as a reference model that portrays these components and categorizes them into different layers.

The Decentralized Identity Foundation (DIF)\footnote{Decentralized Identity Foundation (DIF): \url{https://identity.foundation}} has previously created a reference model that maps the layers of interoperability with the components\cite{interoplayerswhimsical}. Yildiz worked on the consolidation and visualization of these components and layers in the FAQ page of the DIF\cite{dif-faq}. In this tutorial, the reference model is described in detail to understand the differences between technology stacks and, therefore, the interoperability problem.

\section{A Definition of SSI Interoperability - Levels of Interoperability} \label{taxonomy}

In a broad sense, Palfrey et al. define interoperability as the ability to exchange data, and other information between systems, applications, or components \cite{palfrey2012interop}. However, the term interoperability has a slightly different meaning in different verticals. For instance, interoperability in eHealth is defined as the ability of information systems and the embedded devices, and the applications within to access, integrate, and cooperatively use the information \cite{medicalinterop}. In IoT, interoperability can be defined as the ability to exchange and interpret information between things through common standards, protocols, and data exchange formats\cite{konduru2017challenges}. Similarly, in Online Social Network (OSN) services, interoperability is understood as the ability to transmit, interpret, and process information between different users, services, and platforms \cite{goendoer2018seamless}. Although the meanings of interoperability and the means to achieve it are different across the verticals, they use a framework to describe interoperability on different levels. The European Telecommunications Standards Institute (ETSI) accepts this interoperability framework while acknowledging different meanings behind interoperability depending on the vertical\cite{etsiInterop}. These levels of interoperability are:

\begin{itemize}
    \item Technical Interoperability: Interoperability in the technical level refers to the ability to communicate from machine to machine via hardware, software, and other technologies. This kind of interoperability focuses on the underlying protocols and infrastructures that allow them to function. 
    \item Syntactical Interoperability: The concept of syntactical interoperability is usually associated with data formats. Certainly, messages transferred by communication protocols need to be encoded and have syntax, even if the encoding consists of nothing more than bit tables. Data and content are carried over many protocols, and these can be represented using high-level transfer syntaxes like HTML, XML, or ASN.1.
    \item Semantic Interoperability: Semantic interoperability refers to the common interpretation of the content between the sender and receiver.
    \item Organizational Interoperability: Organizational interoperability is the ability of organizations to communicate effectively and exchange (meaningful) data (information), despite using various information systems and operating over different environments. Cross-cultural and cross-regional infrastructures are possible \cite{etsiInterop}.
\end{itemize}

\begin{figure}[!ht]
	\centering
	\includegraphics[width=85mm]{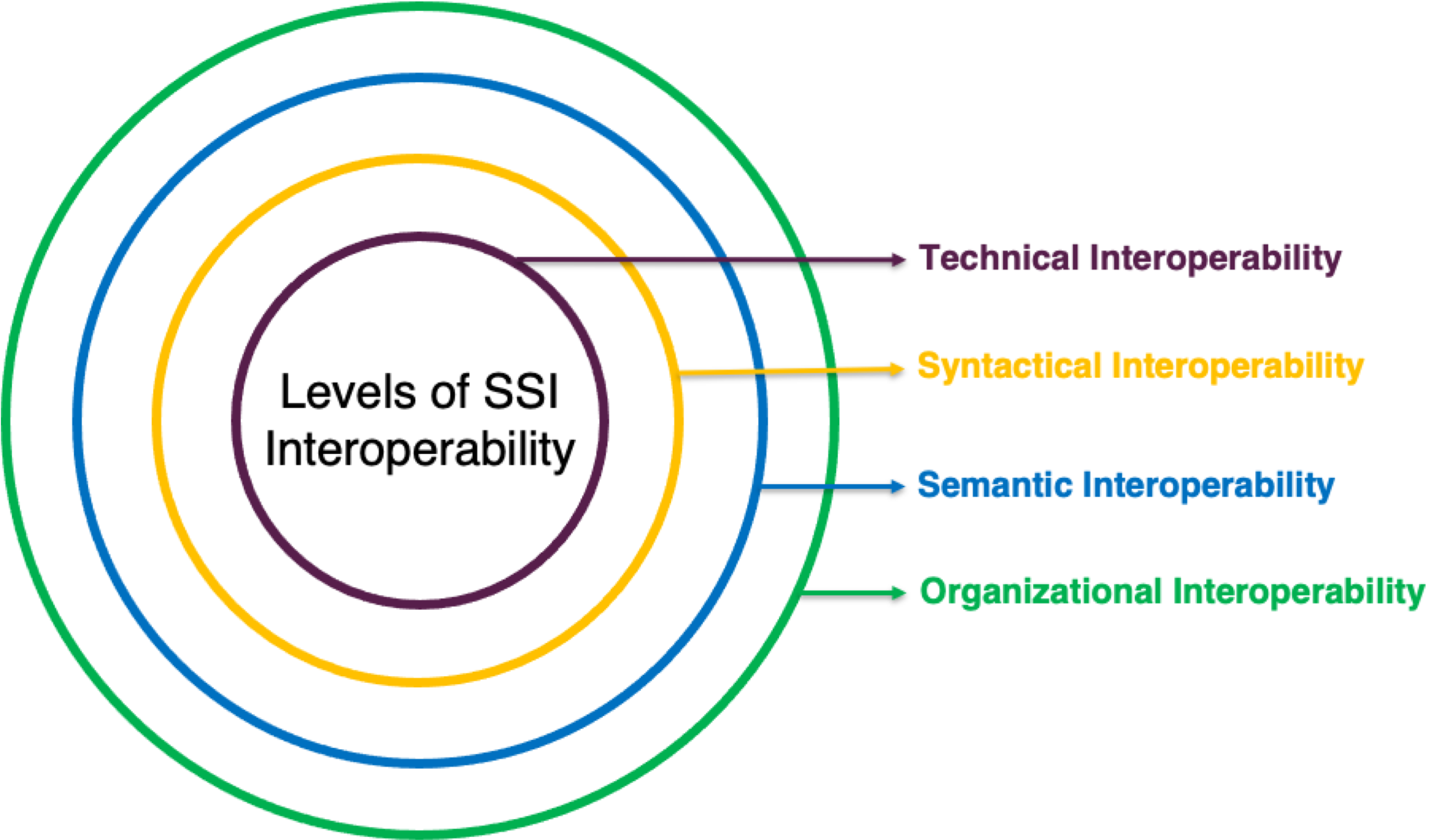}\\
	\caption{Levels of SSI Interoperability, adopted from \cite{etsiInterop}}
	\label{fig:levelsofinterop}
\end{figure}

According to any of these definitions, the higher levels of interoperability require the fulfillment of the lower-level ones. For example, one can only reach semantic interoperability after achieving technical and syntactical interoperability. Figure \ref{fig:levelsofinterop} presents the levels of interoperability and their dependencies with each other. 

To establish a common understanding of the meaning of interoperability in the domain of SSI, a definition needs to be created which is aligned with existing frameworks used by standardization organizations. Thus, we present a definition of SSI interoperability that, according to recognition of ETSI, consists of four levels.

\subsection{Technical Interoperability}
At the fundamental level, technical interoperability covers the communication and information exchange between various software representing the SSI actors or creating trust between actors. To achieve technical interoperability standards for, for example, communication protocols, connection, and exchange of self-sovereign identities must be utilized by the SSI actors. If the used standards cannot be agreed upon, technology stacks must consist of multiple standards, or the SSI actors must agree on a new set of standards and protocols. If none of these options are possible, middleware utilization can help achieve technical interoperability. However, middleware comes with contradicting properties to SSI, such as centralization, that must be considered. 

Interoperability with existing Identity Access Management solutions can be considered mostly as a part of technical interoperability. It is possible to integrate Identity Access Management solutions with self-sovereign identities. SSI-based authentication was proposed by Hong et al. and is compliant with OAuth 2.0 \cite{electronics9081231}. Using the OIDC protocol for authentication, Gruner et al. designed a component-based architecture for integrating uPort\footnote{uPort: \url{https://www.uport.me/}} and Jolocom\footnote{Jolocom: \url{https://jolocom.io/}} into web applications \cite{8935015}. Furthermore, Lux et al. implemented a proof of concept for SSO schemes based on the Hyperledger Aries\footnote{Hyperledger Aries: \url{https://www.hyperledger.org/use/aries}} based technology stack that used OIDC attributes \cite{Lux2020BRAINS}. Lastly, Yildiz et al. proposed a hybrid solution that uses self-sovereign identities for authentication and SAML response to the service provider from the IDP for authorization \cite{Yild2109:Connecting}.

\subsection{Syntactical Interoperability}
According to the proposed definition, data exchange formats are a part of syntactical interoperability. It has the least incompatibilities between SSI actors because most technology stacks use JSON or its extension JSON-LD as a data exchange format in almost all cases. There are other supported data formats, such as CBOR\footnote{Concise Binary Object Representation: \url{https://datatracker.ietf.org/doc/html/rfc7049}} and ASN.1\footnote{Abstract Syntax Notation One: \url{https://www.itu.int/ITU-T/recommendations/rec.aspx?rec=x.680}}. However, these can be easily supported by adding interpreters to a technology stack if needed. Additionally, the data exchange format is embedded in specific components and protocols part of technical interoperability. For instance, JSON is used as the data exchange format when combined with JSON Web Tokens and JWS. 

\subsection{Semantic Interoperability}
Semantic interoperability is the ability to interpret and understand the information exchanged unequivocally between a sender and a receiver. In the SSI domain, the content interpretation is done not by humans but rather by machines, meaning the software deployed by the SSI actors should interpret the exchanged data without any additional effort. Technical and syntactical interoperability is required to achieve semantic interoperability, as exchanging data between parties is the necessary step before interpretation. 

Technology stacks can include existing ontologies, use new ones, or utilize semantic web\cite{villa2017semantics} to achieve semantic interoperability. In the case of SSI, Linked Data and Linked Data Context is the most common way to create semantic disambiguation and achieve semantic interoperability. However, due to the other restrictions and design choices, not every SSI technology stack uses Linked Data. Thus, attaining semantic interoperability stands to be one of the SSI domain's key challenges. 

\subsection{Organizational Interoperability}
Organizational interoperability is the ability to exchange and interpret information between different organizations. In the case of SSI, this is a rich topic since an organization can be a distributed ledger like the Ethereum network, a cooperative that governs a distributed ledger network like Sovrin\footnote{Sovrin: \url{https://sovrin.org/}}, or a service provider outside of the user's ecosystem. For example, identity holders might be a part of the Hyperledger Aries ecosystem and have verifiable credentials issued by an issuing actor using the Aries technology stack. However, they would like to access services from a service provider powered by the uPort stack utilizing the Ethereum network. In the most advanced cases of organizational interoperability, the identity holders would be able to prove their digital identities to the service provider without additional effort from both parties.

There is also economic and legal interoperability as sub-levels within the organizational interoperability. Economic interoperability is the ability to create a value chain between organizations. For example, an organization can enable this level of interoperability by allowing the consumption of credentials by another organization to create added value. It can then issue a new credential or allow for notarization of the credential from the other organization by a trusted accredited organization to increase its level of assurance. However, in some cases, Organization A might not want interoperability with Organization B because it does not fulfill the specific legal requirements necessary for Organization A. For example, Organization B might not be compliant with the Global Data Protection Regulation (GDPR)\footnote{Global Data Protection Regulation: \url{https://gdpr-info.eu/}}, electronic Identification, Authentication and Trust Services (eIDAS)\footnote{eIDAS Regulation: \url{https://digital-strategy.ec.europa.eu/en/policies/eidas-regulation}}, or the Federal Information Security Management Act (FISMA)\footnote{Federal Information Security Modernization Act: \url{https://www.cisa.gov/federal-information-security-modernization-act}}. In that case, Organization A might choose not to exchange and interpret information from the other organization. If they do, this type of interoperability is also referred to as legal interoperability. 

\section{SSI Reference Model}
\label{layers}

The levels of interoperability enable a common understanding based on the recognized and standardized levels. However, it does not enable an understanding of the differences between technology stacks, which is crucial for solving the interoperability problems. In other words, the definition helps organizations set interoperability goals by aiming for a level of interoperability, but it does not help how to reach those goals. To address the question of how a systematic mapping of components of technology stacks is necessary. Therefore, we present a multi-layer reference model that abstracts components and considerations that build up an SSI technology stack.

\begin{figure}[ht!]
	\centering
	\includegraphics[width=85mm]{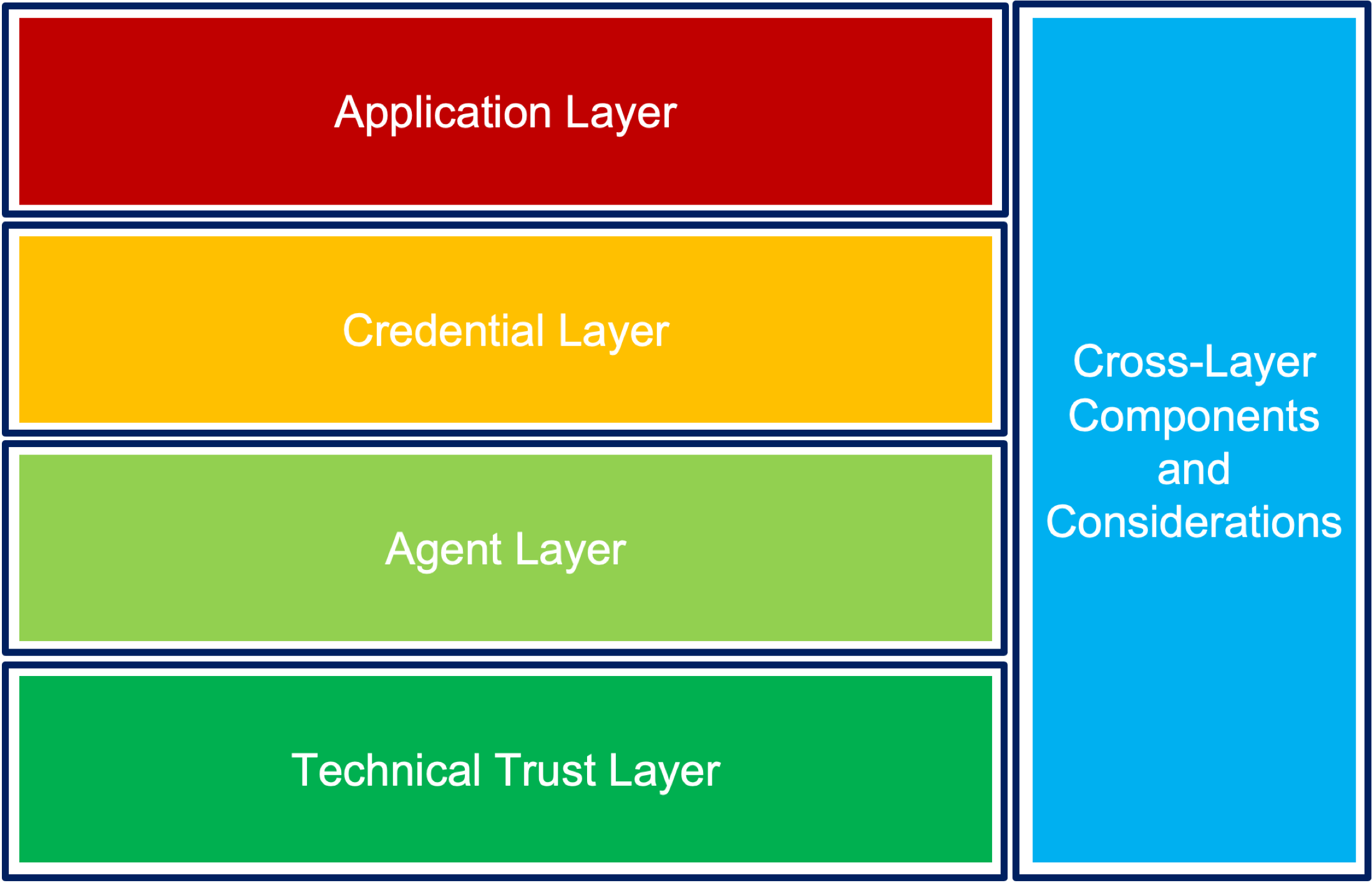}\\
	\caption{SSI reference model, simplified}
	\label{fig:refModel_simple}
\end{figure}

We take the preliminary work of the DIF Interoperability Working Group and describe each layer and component. There are 28 components and considerations that build up an SSI technology stack. This section is structured as follows: In Subsection \ref{sec:trust_layer}, we discuss the components and considerations regarding the public trust layer. Subsection \ref{sec:agent_layer} covers the components related to the agent layer, and Subsection \ref{sec:credential_layer} describes the credential layer. The application layer is discussed in Subsection \ref{sec:app_layer}, and we finalize this section with the discussions regarding components related to multiple layers in Subsection \ref{sec:x-layer}. Figure \ref{fig:refModel_simple} shows a simplified view of the reference model.

\subsection{Technical Trust Layer}
\label{sec:trust_layer}
Like any other digital identity, self-sovereign identities must be issued based on trust that a verifier can cryptographically validate in later stages of the credential life-cycle. Similarly, trust must be created before exchanging information between an identity holder, an issuer, or a verifier. Therefore, as creating trust between SSI actors is the first step in enabling communication and issuance of credentials, the technical trust layer is the first layer of the SSI reference model. 

SSI relies on the concept of \textit{identifiers} to uniquely reference issuers and holders on a worldwide basis. While a broad range of identifier schemes have been created, i.e., the \textit{International Standard Book Number} (ISBN), the \textit{Global Trade Item Number}, or the \textit{Uniform Resource Locator} (URL), SSI makes use of the recently emerging scheme of \textit{Decentralized Identifiers} (DIDs), which is standardized by the W3C \cite{did-core}. Listing \ref{listing:simple-DID} shows a DID example.

\begin{lstlisting}[caption=A simple DID, label=listing:simple-DID, captionpos=b] 
 did:ethr:hw8b234xnola4u80rf426hdwa4
 \_/ \__/ \________________________/
  |    |              |                         
  |   DID       Method-specific
  |  Method       Identifier
  |
Scheme      
\end{lstlisting}

The creation of trust revolves around DIDs that represent the SSI actors. The technical trust layer contains the DID-related components, such as VDRs for their storage and how they are resolved into documents that contain related information. Therefore, it is essential to differentiate between the technical and legal trust, as the former enables the creation of trust between SSI actors with technologies such as public-key cryptography, while the latter empowers SSI actors to trust each other within a legally binding trust framework. 

\subsubsection{DID Document}
A DID follows the syntax of Uniform Resource Identifiers (URI), is globally unique, persistent, and entirely under the control of the entity the DID belongs to, which is called \textit{DID subject}. In addition, each DID is associated with a dedicated \textit{DID document}, see Listing \ref{listing:simple-DID-document} for an example.

\begin{lstlisting}[caption=A simple DID document, label=listing:simple-DID-document,captionpos=b] 
{
  "@context": "https://www.w3.org/ns/did/v1",
  "id": "did:example:1234567890abcdefghi",
  "verificationMethod": [{
    "id": "did:example:1234567890abcdefghi#key-1",
    "type": "Ed25519VerificationKey2020",
    "controller": "id": "did:example:1234567890abcdefghi",
    "publicKeyMultibase": "zH3C2AVvLMv6gmBNam3uVAjZpfkcJZwDwnZn6z3wXmqPV"
  }],
  "service": [{
    "id":"did:example:1234567890abcdefghi#requestIssue",
    "type": "IssuanceService", 
    "serviceEndpoint": "https://example.com/issueCredentials"
  }]
}   
\end{lstlisting}

As can be seen, by this example, a DID document includes the DID of the DID subject and, optionally, the DID of a \textit{DID controller}, which is an entity having the right to modify the document if the DID subject itself does not do this. Furthermore, it defines the \textit{verification material} required by entities interacting with the DID subject for different purposes like verifying digital signatures as needed for SSI, for authentication, or the exchange of confidential messages. The verification material is composed of one or several \texttt{verificationMethod}, which can be used on a general basis or bound to one of the purposes above. A verification method also has a unique identifier for referring to it and defines a type of asymmetric encryption and an associated public key. Optionally, the controller refers to the entity that owns the associated private key if it is not the DID subject. Finally, the optional \texttt{service} property describes how the DID subject can be reached to access a service such as credential issuance.

An organization that wants to be found publicly creates public DIDs. Their DID documents can be stored directly in a VDR such as distributed ledger, generated on-demand from raw data stored on a distributed ledger, or in distributed file storage systems such as IPFS. DID documents stored this way can create the necessary technical trust for actors to interact with the DID subject since the DID document can only be manipulated if the entity has the private key to the public key declared in the verification method of the DID document. Furthermore, as the DID documents belonging to public DIDs are immutable and highly available, they can be looked up by an actor with confidence that the information related to a DID in its DID document is not tampered with by a third party. This is the essence of creating technical trust between SSI actors, which then enables the interactions that occur on the higher layers.

\subsubsection{DID Methods}
Since a DID is only an identifier, it does not have a special meaning on its own. It is the DID document that contains the necessary information to establish trust between the SSI actors. In order to resolve into a DID document, a DID needs instructions from its DID method. DID method-specific instructions (operations) are at least the following and are called CRUD operations\cite{fielding2009post}:

\begin{itemize}
\item \texttt{Create}: Operation to create a new
DID, set authentication keys, and define a service endpoint.
\item \texttt{Read (Resolve)}: Operation to
resolve the DID into the DID document containing the key information.
\item \texttt{Update}: Operation to update a DID, 
set or extend authentication keys, define a service endpoint, or rotate
authentication keys for the DID.
\item \texttt{Delete (Revoke)}: Operation to
revoke the DID to restrain its usage in the latest state of the
verifiable data registry. Deleting a public DID stored on a VDR is not possible with conventional means
since the older states of the verifiable data registry are traceable.
Delete operation merely revokes the DID so that it cannot be used in the
future.
\end{itemize}

\subsubsection{DID Resolution}
DID resolution is one of the four essential operations mentioned above. It contains the read operations that allow a DID to resolve into a DID document. On top of the DID resolution, the DID URL dereferencing runs as a process responsible for retrieving a DID URL’s representation. Software and or hardware that can run both processes is called a DID resolver\cite{didresolution}.

Dereferencing is also explained in the DID-Core document by W3C\cite{did-dereferencing}. It is a function that dereferences a DID URL into a resource with contents based on the components of the DID URL. It includes the DID method, method-specific identifier, query,  path, and a fragment. The process of the URL dereferencing is highly dependent on the DID resolution contained in the DID URL. The function returns the final resource after going through the steps mentioned above. A resource can be anything that a URI can identify. In some cases, the DID subject might be the resource identified by a DID\cite{did-core}.

\paragraph{Resolution of Pairwise DIDs}
So far, we have discussed the resolution of public DIDs. However, to ensure privacy, DIDs of identity holders must not be written in a VDR. Currently, most VDRs are DLT-based, and it is practically not feasible to delete a record if the identity holders invoke their right to be forgotten according to GDPR\cite{righttobeforgotten}. Moreover, from the regulator's point of view, the end user must consent for their data to be processed by a known processor. Since anyone can process the data in public DLTs, the consent required by GDPR is not practical to implement as the user consents to their data to be processed by unknown entities. Instead, identity holders are represented by a type of DIDs called pairwise DIDs. If the analogy to public DIDs in PKI is an X.509 certificate that a certificate authority (CA) issues, the pairwise DIDs are the equivalent of self-signed X.509 certificates commonly found in end-user devices. 

Pairwise DIDs require a different type of DID method specification, which describes the deployment of DIDs in situations where a pairwise or an N-wise relation is needed. It includes the representation of SSI actors, such as identity holders, without a persistent identifier for the interactions with different actors. For instance, Mike wants to communicate with his university. For that purpose, he is represented with a pairwise DID A. If Mike wants to communicate with another organization, such as a bank, he is represented with a pairwise DID B. This approach is not required for legal entities, and they can be represented by a static public DID for interactions with various actors. 

The method specifying the pairwise and N-wise DIDs is called \texttt{did:peer:} and the DIDs from this method come with the following benefits\cite{didpeer}:

\begin{itemize}
\item they have no transaction costs. CRUD operations are executed within the software of the SSI actor,
\item they can scale as much as the user needs,
\item concerns regarding privacy are much reduced,
\item they are not stored in a VDR
future, and
\item \texttt{did:peer:} can be integrated into any other DID method by grafting, meaning they can be resolved by the other DID method.
\end{itemize}

Grafting can be done in different ways. The most common three ways are\cite{didpeer}:
\begin{itemize}
\item combining Base64 Body of peer DID with another DID method prefix,
\item extending the DID method prefix with the peer did, e.g., \texttt{did:sov:peer:xyz}, and
\item by re-encoding the base64 body of the peer DID with the ruleset of another DID method.
\end{itemize}

Since pairwise DIDs are not stored in a VDR, the root of trust for those DIDs is the entropy in the initial public key. Therefore, one public key cannot be used to create multiple pairwise DIDs\cite{didpeer}. 

\subsubsection{DID History}
Depending on the chosen technology and type of DID, a DID can have a history or be current-only state. A DID is in current-only state when it is not anchored to a VDR. The most prominent current-only state DIDs are peer DIDs. Public DIDs, on the other hand, have queryable historical states, meaning the history of a DID can be traced if it is anchored to a VDR such as a distributed ledger. Moreover, if a DID supports certain functionalities such as key rotation, the DID counts as a revocable DID. Finally, if a public DID is written in a distributed ledger, it is impossible to delete the entry without rolling the ledger state back. For this purpose, a DID can be denied access, meaning that no DID document is returned when the DID is resolved. This method is called tombstoning and counts as one of the tools in case sensitive information is written in a distributed ledger.

\subsubsection{DID Anchor Types}
Public DIDs and corresponding DID documents must be stored in a VDR to create the technical trust that enables interactions between SSI actors. For instance, the Bitcoin network can be utilized for resolving DIDs with the \texttt{did:btcr:} method. The reasons for utilizing distributed ledger technology (DLT) are its verifiability, availability, and immutability properties. These properties fit the requirements of public DIDs and DID documents for creating technical trust. However, distributed ledgers are not the only options for DID anchors. Non-distributed ledger approaches, such as the Key Event Receipt Infrastructure (KERI)\footnote{Key Event Receipt Infrastructure (KERI): \url{https://keri.one/}}, web servers, Interplanetary Filesystem (IPFS)\footnote{Interplanetary Filesystem (IPFS): \url{https://ipfs.io/}}, and even Public Key Infrastructures (PKIs), can serve as DID and DID document anchors. In some cases, the trust is in the technology and the institution running the instances. In conclusion, there are two archetypes of anchor types: DLT-based anchors such as the Ethereum Network and non-DLT-based anchors such as KERI or PKI.

\subsubsection{DID Scaling}
Distributed ledgers such as Ethereum or Bitcoin can be slow and expensive for executing CRUD operations on DID documents. Since these public permissionless distributed ledgers have a limited amount of transactions per second on layer 1, layer 2 (off-chain\cite{layer2scaling}) scaling options can be taken to enable more efficient CRUD operations. As a layer 2 scaling option, KERI creates similar styles of the chain of trust known in the PKI and implements them in a ledger-agnostic manner\cite{KERI}. The security and immutability properties of the KERI implementation are based on the trust anchor it is connected with. Similarly, \textit{Sidetree} is a layer 2 scalability option that is ledger-agnostic. The DIF maintains the project, mainly developed by Microsoft for its Identity Overlay Network (ION). The main focus is to create a layer 2 solution on top of the Bitcoin network\cite{microsoftion}. 

The DID scaling is only relevant if the SSI stack is not using a specifically designed identity network. Ecosystems such as the IDunion\footnote{IDunion: \url{https://idunion.org/?lang=en}} or the eSSIF\footnote{European Self-Sovereign Identity Framework (eSSIF): \url{https://ec.europa.eu/cefdigital/wiki/pages/viewpage.action?pageId=379913698}} do have designated identity networks. Therefore, they can write with much higher transactions per second with low cost to their distributed ledgers than public permissionless ledgers.

\subsubsection{DID Anchored Services}
Services that are enabled via DIDs and DID documents are called anchored services. One such service is identity hubs, a data storage and message relay system that can be used to locate public or permissioned data related to a DID\cite{identityhub}. Another service is the encrypted data vaults (EDVs) which can store any data in cloud storage or distributed ledger that can only be accessed via
authentication keys described in a DID document\cite{edv}. The advantage of EDV is that unless the private keys are compromised, there is no risk of data theft because of a cloud service hack since the data is encrypted.


\subsection{Agent Layer}
\label{sec:agent_layer}
In the SSI paradigm, actors such as issuer, holder, and verifier, are represented by agents, which then are represented by DIDs. Through the technical trust layer components, these DIDs resolve into DID documents containing public keys and service endpoints related to them\footnote{see Listing \ref{listing:simple-DID-document}}. With this information, an SSI agent can sign, encrypt, and forward messages related to credentials and agent-to-agent communications, including routing and forwarding messages. An SSI agent also stores the keys to perform these activities, and depending on whether it is an edge or cloud agent, it can store credentials. A wallet is the storage medium for the keys within an agent software. Edge agents are agents running in the domain of the identity holder, while cloud agents stereotypically are the agents representing issuers and verifiers. This subsection explains the components related to the SSI agents, which builds the second layer of the reference model.

\subsubsection{Envelope}
One fundamental role of an agent is to communicate with other agents. Since agents are represented by DIDs, they can be reached by an endpoint mentioned in their corresponding DID documents. The messages can be encrypted for the recipient's public key, also found in its DID document. An envelope is a component specifically responsible for exchanging messages between agents in a secure manner. According to the subject of the information security, used in various domains, such as cryptography and PKI, specific properties make communication, and the exchanged information secure\cite{infosec}:
\begin{itemize}
\item \textit{Authenticity} is the property that lets the receiver know that the message is encrypted with the private key of the sender's public/private key pair.
\item \textit{Integrity} means the recipient will notice if the message is tampered with by an unauthorized third party. This property makes sure that the recipients know if the message they received is tampered with or intact.
\item \textit{Confidentiality} ensures that the message is only intended for the	authorized view. In public-key cryptography, this is done by encrypting the message by the recipient's public key so that only the recipient's private key can the message be read.
\item \textit{Non-Repudiation} means that the delivered message's authenticity and origins cannot be disputed. It requires authenticity and integrity to make a message non-repudiable.
\item Additionally, in communications with the involvement of DIDs and DID documents, \textit{routing} is the process of finding a route to deliver information. This property enables the \textit{availability} of a route between two agents.
\end{itemize}

Fulfilling these properties, DIDComm is a draft for interoperable communication between any two entities that DIDs represent. It is also the most adopted message envelope in the technology stacks. The draft realizes a decentralized way of addressing and relaying messages between entities by creating an authenticated communication method via which the entities can exchange data. 
On a protocol level, DIDComm follows a message-based and asynchronous approach, with individual messages being delivered to a receiver in a simplex fashion.
Messages facilitate mutual authentication via DIDs, allowing a receiving party to verify a message's integrity and authenticity easily.
To do so, the receiving party must resolve the sender's DID to the respective DID document, which specifies the public key used with DIDComm.

DIDComm messages are formatted using JSON Web Messages (JWM) \cite{looker2020jwm}. The JWM specifies message header fields, such as a pseudo-unique message ID, message type, and the sender's and receiver's DIDs.
By resolving the respective DIDs to a DIDComm endpoint, messages can be routed to their destination and answered if necessary.
DIDComm supports three types of messages: simple plaintext, signed, and encrypted messages:

\begin{itemize}
\item DIDComm Plaintext Messages: Plaintext messages without any protective envelope.
    Plaintext messages lack support for verifying confidentiality and integrity and should therefore not be used for communication across domains.
    The message type can be used for purposes that do not require any security, such as debugging.
	
\item DIDComm Signed Message: JWM-formatted message exchange between two or multiple participants. 
	DIDComm signed messages may be used in scenarios where the message contents are non-confidential and only integrity, authenticity, and non-repudiation need to be ensured.

    \item DIDComm Encrypted Message: Encrypted messages further encrypt the JWM using the encryption keys specified in the receiver's DID document.
\end{itemize}

Although there is another message envelope as a part of the OIDC technology stack (see Section \ref{considerations} Subsection \ref{agent_Interop_Considerations}), DIDComm is the predominant message envelope found in technology stacks.

\subsubsection{Transport}
The message in the envelope must be transported with a medium to reach the recipient. DIDComm is transport-agnostic, meaning any medium can transport the envelopes. In the following, we discuss the possible transport protocols implemented in SSI technology stacks without being exhaustive.

The Hypertext Transfer Protocol (HTTP) is a protocol for exchanging hypermedia documents such as HTML. HTTP is a stateless protocol that does not keep any data between two separate requests\cite{mozilla-http}. HTTP's typed and negotiated data representation allows for independent development of systems regardless of the data included in the transfer\cite{http-ietf}.

Another widely adopted protocol is WebSocket, which enables two-way communication between an untrusted client running in a regulated environment and a remote host whose communications have been approved by the client. Typically, web browsers use an origin-based security model for this. Over TCP, the protocol is composed of a handshake followed by basic message framing. WebSocket protocol aims to provide browser-based applications that require two-way communication with servers with a mechanism that does not require the opening of multiple HTTP connections\cite{websockets-ietf}.

Besides HTTP and WebSocket, any other transport technology, such as the NFC and Bluetooth can be deployed. The message envelopes require corresponding public key information known to agents to exchange data securely. For the initial exchange, when this information is unknown, the agents can initiate communication via the out-of-band protocol\cite{ariesOOB}, with which the necessary information to establish a secure communication is shared without encryption. This information can be shared in a medium that is only reachable via HTTPS and TLS to ensure a level of trust.

\subsubsection{Control Recovery}
Every transaction, from creating a communication channel between two parties with DIDComm to the exchange of self-sovereign identities, requires possessing and controlling DIDs and corresponding keys. Since no central authorities can restore user access, a backup and restore functionality is required in case of users lose access to their keys. Control recovery contains the essential components for recovering these keys in the case of device loss or migration. 

There are mainly two methods of control recovery in a decentralized environment. The first involves the creation of deterministic keys based on a mnemonic code called a seed phrase. It allows the creation of many keys by deriving them from a master key\cite{SeedPhraseshaik2020securing}. The seed phrase provides access to the master key and its derivations and is the industry standard for cryptocurrencies with the implementation of the Bitcoin Improvement Protocol 39 (BIP-39)\cite{bip39}. SSI technology stacks take the same approach for key recovery. 

The second method is a system built on deterministic keys, and seed phrases called the decentralized key management system (DKMS). DKMS describes best practices for using and rotating keys, recovery methods, multi-device management, and key generation. DKMS also comes as a protocol that can be implemented in an SSI technology stack\cite{DKMS}. In addition, DKMS describes recovery types such as offline or social recovery types. For example, Shamir Secret Sharing allows the division of the seed phrase and recovery with n out of N parts of the seed phrase\cite{shamir}. These recovery types can be used as-is or with the help of applications like SeedQuest, FuzzyVault, or the Horcrux Protocol, which help recover seed phrases by playing a game\cite{seedquest}, use of biometric data\cite{fuzzy}, and instantiating Shamir Secret Sharing\cite{horcrux} respectively.

\subsubsection{Key Operations}
Key operations is a broader term for utilizing public/private key pairs. For example, as mentioned earlier, creating a message envelope requires public/private key pairs. For instance, the sender signs the message with their private key (authenticity) for the public key of the receiver, which can only be decrypted by the receiver (confidentiality)\cite{infosec}. 

Additionally, an issuer creates self-sovereign identities by signing a set of claims to make these claims cryptographically verifiable. When presenting to the verifier, the holder agent also creates a cryptographic proof of ownership that the shown credential is bound to the identity holder. These activities, which we will cover in Section \ref{sec:credential_layer}, also require public/private key pairs. Therefore, they are also a part of key operations.

\subsubsection{Data Portability}

Portability is the sixth principle written by Christopher Allen in 2016, mentioning the necessity of transportability of digital identities to different mediums. According to Allen, self-sovereign identities should not be held exclusively by a centralized entity \cite{allen2016path}. As the new identity paradigm comes with new infrastructure, it is no surprise that SSI is challenging the existing identity paradigms and the role of identity providers. In the current identity paradigms, identities are not portable, and recalling service from an identity provider results in loss of identity and access to the related services\cite{Yild2109:Connecting}. Data portability is the component that enables the migration of self-sovereign identities and their related keys to another agent. Data portability involves recovering or transferring keys using DKMS methods like control recovery. In addition to control recovery,  data portability enables the transfer of self-sovereign identities to another agent within the user’s domain. 

The Hyperledger ecosystem with the Aries .NET Framework makes it possible to export issued credentials to another wallet. It means the credentials are exported along with the DIDs, keys, and credential binding information. The import/export function of the Indy-SDK enables this feature. Keys and DIDs are exported using the BIP-39 Standard\cite{walletportsovrin}. 

Another development outside the Hyperledger ecosystem is the Universal Wallet, a specification maintained by the W3C. It is not a change to the existing wallet architectures. Instead, it proposes a specification to promote wallet portability, including verifiable credentials and DIDs\cite{universalwallet}. Universal Wallet aims to encourage wallet portability by creating a data model for every wallet and agent type related to SSI by modeling the wallet types and different functionalities of a wallets\cite{uniwalletspec}. This way, various wallet providers could translate wallet data and import it into their architecture. 

In the current landscape of SSI, data portability is seen as a nice-to-have functionality. Moreover, from a regulatory point of view, some credentials, such as eID, must not be portable. Currently, there are efforts in the Wallet Security Working Group at the DIF, working on a hybrid approach to enable data portability for the low level of assurance credentials while disabling it for a high level of assurance credentials. This hybrid approach is called differential credentialing.

\subsection{Credential Layer}
\label{sec:credential_layer}
The technical trust and agent layers create the necessary fundamentals for the interactions between SSI actors through their agents. Once the technical trust is established and agents can communicate with each other securely, they can issue, store, or verify self-sovereign identities. The layer that contains the components related to the types, proofs, revocation, exchange, and binding of self-sovereign identities is called the credential layer.

\subsubsection{Credential Format and Proof}
A self-sovereign identity is an object containing, among other things, a set of key/value pairs called claims. Claims represent statements about a subject, an identity holder, expressed by subject-property-value relationships. Listing \ref{listing:simple-claim} shows a simple example.

\begin{lstlisting}[caption=A simple claim, label=listing:simple-claim,captionpos=b] 
"credentialSubject": {
    "id": "did:example:1234567890abcdefghi",
    "name": "Mike",
    "enrolledStudentAt": "TU Berlin",
    "subject": "Computer Science"
}
\end{lstlisting}

In this example, Mike is the identity holder. As seen in the listing, the identity holder is referenced by its DID followed by the property-value pair. The properties in this example are \texttt{name}, \texttt{enrolledStudentAt}, and \texttt{subject}, which have the values of Mike, TU Berlin, and Computer Science, respectively. These claims can be written or tampered with by any party. To make them tamper-resistant and verifiable, they must be contained in an environment that enables these properties. Two of the widely adopted containers are verifiable credentials and verifiable presentations:

A \textit{verifiable credential} (VC) is a tamper-resistant container including one or multiple claims about an identity holder. It is created by an issuer, who creates or takes the respective claims, checks them for correctness, and digitally signs them. Listing \ref{listing:simple-verifiable-credential} shows an example.

\begin{lstlisting}[caption=A simple verifiable credential, label=listing:simple-verifiable-credential,captionpos=b]
"id": "http://example.edu/credentials/123",
"type": ["VerifiableCredential", "StudentCredential"],
"issuer": "did:example:rstuvwxyz0987654321",
"issuanceDate": "2021-11-23T18:00:00Z",
"credentialSubject": {
    "id": "did:example:1234567890abcdefghi",
    "name": "Mike",
    "enrolledStudentAt": "TU Berlin",
    "subject": "Computer Science"
},
"proof": {
    "type": "Ed25519Signature2020",
    "created": "2021-11-23T18:00:00Z",
    "proofPurpose": "assertionMethod",
    "verificationMethod": "did:example:rstuvwxyz0987654321#key-1",
    "jws": "pYw8XNi1..Cky6Ed="
}
\end{lstlisting}

The ingredients of a VC are the claims, metadata, and proof. The metadata includes an identifier for the VC, the DID of the issuer, and the credential's type and issuance date. The type field value is usually a list of keywords, each referring to the name of a well-defined or standardized data scheme the claims in the verifiable credential follow. The metadata guarantees semantic interoperability at a high level between verifier and issuer. Without this specification, the verifier agent cannot interpret and verify the credential.

The proof makes the VC tamper-resistant by using cryptographic methods like digital signatures. Such proof guarantees the authenticity of the claims; that is, an issuer has confirmed the claims by signing them along with the metadata with a private key of a public/private key pair associated with the issuer DID. The W3C does not prescribe or restrict the range of cryptographic methods used for proofs but identifies, i.e., data integrity proof \cite{longley2018dataintegrity} as a potential candidate. For example, Listing \ref{listing:simple-verifiable-credential} makes use of a data integrity proof. The value of the proof field contains the value of the proof, that is, the digital signature as a JSON web signature (JWS) using Edwards-Curve Digital Signature Algorithm\footnote{Edwards-Curve Digital Signature Algorithm: \url{https://datatracker.ietf.org/doc/html/rfc8032}}, and meta-data needed for verifying the proof. The structure of the metadata depends on the representation language used. The example consists of a reference to a verification method in a DID document (which provides access to a public key, which is needed for verification), the type of the signature, the signing date, and the purpose of the proof. The latter indicates to the verifier for which purpose the proof has been created. In the example, the value \texttt{assertionMethod} indicates that the verification of this proof provides knowledge of whether the issuer has confirmed the claims (assertion) contained in the VC or not.

The VC compiled by an issuer is handed over to the identity holder, who can maintain it at his preferred location, for example, an agent on his smartphone.

A \textit{verifiable presentation} (VP) is made from one, in some cases, more verifiable credentials when the identity holders need to prove (parts) of their identity to a verifier. In our example, Mike would like to enjoy a discounted ticket to the theater, and for this purpose, he must prove at the box office, that is, the verifier, that he is still a student. To accomplish this, he takes the VC issued by his university and generates a verifiable presentation from it, shown in Listing \ref{listing:simple-verifiable-presentation}.

\begin{lstlisting}[caption=A simple verifiable presentation, label=listing:simple-verifiable-presentation,captionpos=b] 
"@context": "https://www.w3.org/2018/credentials/v1",
"type": ["VerifiablePresentation", "StudentPresentation"],
"verifiableCredential": [{
    "id": "http://example.edu/credentials/123",
    "type": ["VerifiableCredential", "StudentCredential"],
    "issuer": "did:example:rstuvwxyz0987654321",
    "issuanceDate": "2021-11-23T18:00:00Z",
    "credentialSubject": {
        "id": "did:example:1234567890abcdefghi",
        "name": "Mike",
        "enrolledStudentAt": "TU Berlin",
        "subject": "Computer Science"
    },
    "proof": {
        "type": "Ed25519Signature2020",
        "created": "2021-11-23T18:00:00Z",
        "proofPurpose": "assertionMethod",
        "verificationMethod": "did:example:rstuvwxyz0987654321#key-1",
        "jws": "pYw8XNi1..Cky6Ed="
    }
}],
"proof": {
    "type": "RsaSignature2018",
    "created": "2021-11-30T14:00:00T",
    "proofPurpose": "authentication",
    "verificationMethod": "did:example:1234567890abcdefghi#key-1",
    "challenge": "1f44d55f-f161-4938-a659o-f8026467f126",
    "domain": "4jt78h47fh47",
    "jws": "BavEll0/I1..W3JT24="
}
\end{lstlisting}

As seen in the listing, a VP wraps the VC and attaches another proof to it. This proof must be compiled when the verifier prompts Mike to send the presentation. The purpose of this proof is to authenticate Mike and avoid an entity that has previously received the verifiable presentation performing a replay attack. To prevent such attack, the presentation contains a \texttt{challenge} that the verifier has randomly generated for inclusion into the proof before signing.  

\subsubsection{Credential Revocation}
To ensure the identity holder may use a credential to access services for a limited amount of time, may it be due to regulations, security, or requirements specific to use cases, a VC may have an expiration date. However, it might be necessary to revoke a VC before its expiration date in specific use cases or merely revoke a credential without an expiration date. This feature is called credential revocation. A revoked credential is not removed from the wallet of the identity holder. Instead, additional cryptographic validation is done by the verifier. Even if the credential is removed, the verifier still validates that the credentials were issued and were valid until a specific time. 

The idea of credential revocation has its roots in public-key cryptography and PKI. For example, the X.509 certificates, the standard format of public-key certificates, are sometimes required to be revoked. There are various reasons for revoking an X.509 certificate, including compromised encryption keys, errors occurring in the issued certificate, and the certificate owner being no longer trusted\cite{x509rev-ibm}. The Certificate Revocation List (CRL) and the Online Certificate Status Protocol (OCSP) are two of the most used revocation mechanisms. CRL is a timestamped list of entries identifying the revoked certificates either signed by a CA or a CRL issuer. The CRL is made available through a public repository. The CRL is updated in regular intervals, varying between mere hours to weeks\cite{crl}. 

OCSP, on the other hand, can be used instead as a supplement to CRL. OCSP enables more timely revocation information than CRL. OCSP clients send status requests to their responders and suspend acceptance of certificates in question until the responders respond. An OCSP response can be in a "good", "revoked", or "unknown" state. If there are no implications of revocation for the certificate serial number, the "good" state is sent. There is no guarantee that the certificate was ever issued or that the time when the response was produced was within the certificate's validity period. Additional requests can be made via response extensions. The "revoke" state indicates that the certificate with the given number has been revoked permanently or temporarily. It can also indicate no records of issuance of the certificate in question using any current or previous issuing keys. "Unknown" means that the responder does not know about the requested certificate, typically due to an unrecognized issuer not served by this responder\cite{ocsp}.

Following similar approaches taken for X.509 certificates in PKIs, some revocation mechanisms for VCs have revocation status lists similar to CRL that can be looked up during the verification. Revocation is an optional component that might be required for specific use cases, and there is more than one implementation method.

\subsubsection{Credential Exchange}
Self-sovereign identities are exchanged securely as messages in envelopes such as DIDComm with transport mediums such as HTTP. However, more than one interaction occurs during the exchange between the agents. These interactions are standardized in exchange protocols. Exchange protocols can be categorized into two types:

\paragraph{Credential Issuance}
Exchange protocols related to credential issuance define the interactions between an issuer and a holder agent during the issuance process. For instance, Mike can negotiate what kind of VC he wants from TU Berlin by stating in a message which claims he wants to have in the VC. TU Berlin can, in return, propose a VC to Mike, and Mike can either accept or decline it. Once accepted, TU Berlin compiles the VC and sends it in a message (encrypted with a message envelope). These interactions are standardized in credential exchange protocols related to the credential issuance process. 

\paragraph{Proof Presentation}
Similarly, exchange protocols related to VPs define the interactions between a holder and a verifier. Following the same instance, Mike goes to a theater and wants to buy a discounted ticket. The theater allows discounted tickets for students, unemployed citizens, and pensioners. Therefore, the verifier agent of the theater asks one for one of the three possible presentations: a student card, an unemployment status document, or a pensioner card. Mike's agent communicates that it can show proof of student card from TU Berlin. The verifier agent agrees to accept proof from a student card created by TU Berlin. Mike presents the VP, and the verifier agent validates the credential and allows the purchase of a discounted ticket. These interactions are standardized in credential exchange protocols related to the proof presentation process.

\subsubsection{Credential Binding}
There are multiple types of digital credentials. For instance, if the subject of the credential is not present, the type of credential is called a bearer credential or a bearer token. An example of this type of credential is a movie ticket: A movie ticket is not bound to a person, but merely a proof of possession \cite{vc-data-model}. Self-sovereign identities addresses use cases where the proof of possession is insufficient.

When the identity holders show VPs, they also need to prove that the underlying VCs were bound to them. For instance, Mike has given a DID to TU Berlin for creating his student card, which the issuer writes in the id field in the credentialSubject are of the VC (see Listing \ref{listing:simple-verifiable-credential}). Mike then proves that the VC is bound to him by creating a VP and signing it with one of the private keys of a public/private keypair related to the DID written in the credentialSubject. This process is an example of a credential binding mechanism.
       
\subsubsection{Credential Delegation}
The topic of credential delegation, which explores the options of chaining credentials to create a chain of trust similar to X.509 certificates, has not been considered as a component since there are special interest groups and publications around it but no current implementations or architectures. Nevertheless, it is an important topic that is worth mentioning. As of May 2022, there are two proof of concepts for credential delegation. One of them is a set of conventions that permit data in a verifiable credential to be traced back to its source while retaining its verifiable quality\cite{ariescredentialchaining}. The other one is mathematical proof that the credential chaining can be done using the self-sovereign identities implemented in the Hyperledger Indy technology stack\cite{camenisch2017practical}. 

\subsection{Application Layer}
\label{sec:app_layer}
We have discussed the first three layers and described the SSI components, which enable the discovery and availability of SSI actors, secure communication between various SSI agents, and exchange and validation of self-sovereign identities. Building on top of the first three layers, the application layer contains the components and business logic related to the use cases. For instance, an issuer application that only issues organizational credentials has a different business logic than a holder application for natural persons. This layer is also considered the fundamental layer for semantic interoperability as it contains the components related to data definitions and vertical considerations specific to use cases. Furthermore, the application layer is also the highest layer in the reference model. 

\subsubsection{Apps and dApps}
Applications utilize the underlying layers to enable services related to use cases. Applications can be complex controllers that contain business logic. They can also be smaller in scale and serve as enablers for other use cases. One such application was created in 2019 to test the connectivity, responsiveness, and security of a channel created by pairwise DIDs using the DIDComm envelope called Trust Ping. Trust in the name comes from the fact that both the sender and receiver can check how secure the connection is by seeing the envelope, algorithms, and keys. Trust Ping can also route these messages to any other agents and relays within the receiver's domain, and the message can be traced using a general message tracing mechanism\cite{trustping}. 

Furthermore, Sabadello et al. created further applications to allow the issuance and verification of credentials with a web-based technology stack. The Universal Issuer is a service that enables the issuance of JSON-LD and JSON credentials with LD and JWT-Proofs, respectively. It uses VC over HTTP transport and can be used by various DIDs from diverse DID methods \cite{uniissuer}. Sabadello et al. have also built the Universal Verifier, which can verify JSON-LD and JSON credentials with data integrity and JWT proofs. It uses CHAPI for credential exchange purposes\cite{univerifier}. These applications are still in progress and do not offer certain functionalities such as revocation or issuance and validation of BBS+ signatures.

Although not popular at the current stage, decentralized applications (dApps) that contain the business logic on smart contract platforms such as the Ethereum network can be created. For example, dApps can enable granting access to services such as decentralized exchanges with legal credentials. 

\subsubsection{Semantic Data Definition}
Along with Linked Data context, semantic data definition plays a vital role in creating semantic disambiguity for credentials and their attributes. Semantic data definition ensures that the claims in a credential are structured in an outsourced domain, in most cases linked by Linked Data context. There are many domains for this purpose, but one of the most prominent places for storing such structured data is schema.org, where structured data can be stored and maintained by community activity. This structured data can be added as an ontology via RDFs, or in the case of SSI, JSON-LD\cite{schemaorg}. 

Similar data structures are possible using different mediums. For example, Comprehensive Learner Record specifies which claims are used for which purposes for degrees, certifications, courses, and competencies\cite{learnerrecord}. Open Badges define the type badge and create semantic understanding between exchanging parties for any learning and achievement credentials\cite{openbadges}. ActivityPub, although in itself a protocol, uses schemas for structuring data according to person type, which suits the use of decentralized social media \cite{activitypub}. These different semantic definitions have one thing in common: They belong to specific verticals.

\subsubsection{Verticals}
Industries have requirements that are specific to the use cases within those industries. Therefore, setting interoperability goals while keeping these requirements in mind is essential. For instance, in the education vertical, Europass Learning Model is a crucial standardization that acts as a bridge between the ELMO/EMREX standard and SSI-based credentials in Europe\cite{europasslearningmodel}, where ELMO/EMREX is seen as the industry standard for exchanging student information in Europe\cite{emrexhandbook}. Similarly, the International Standard Classification of Education (ISCED) plays an equally important role in categorizing, among other things, education levels and EU countries\cite{isced}.

In eHealth, cross-domain data exchange must fulfill specific existing standards and protocols. One of these standards is Fast Healthcare Interoperability Resources (FHIR), a solution built from modular components (Resources) that can be assembled into systems, which then can be used to solve real-world health administration problems. For example, FHIR can be used for exchanging data via mobile phone applications, cloud communication, and server communication between large institutional healthcare providers\cite{fhir}. In addition, for the authorization and authentication of patient records, Patients can use UMA to authorize health clinics to store their data or exchange them with other health clinics\cite{uma}.

There are many other vertical considerations within the sectors mentioned above and any other vertical for which an SSI solution is built. It is not in the scope of this paper to explain every existing vertical and the used standards and protocols for them. However, it is crucial to consider the relevant standards and protocols while designing SSI solutions, building use cases, and setting interoperability goals. 

\subsection{Cross-layer Considerations}
\label{sec:x-layer}
\begin{figure*}[t]
    \centering
    \includegraphics[width=\textwidth]{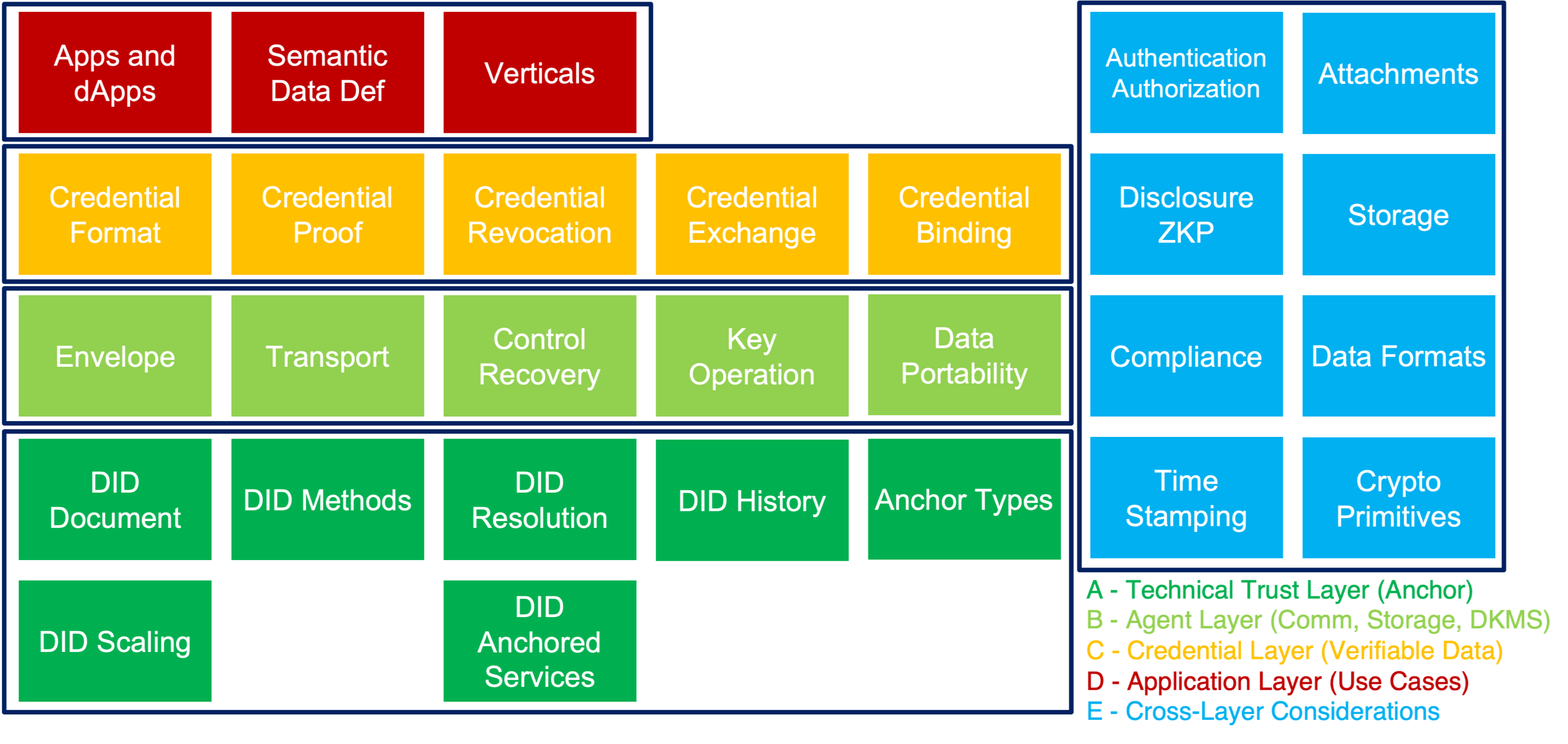}
    \caption{SSI reference model, complete picture}
    \label{fig:layers}
\end{figure*}

Cross-layer considerations are components and regulations that affect more than one component within the four layers. For instance, crypto primitives can enable updating DID documents (a technical trust layer component) and signing verifiable credentials and presentations (credential layer components). It is important to note that different standards and protocols within a cross-layer consideration might be used in components from different layers. For instance, Cryptographic Algorithm A can be the underlying primitive for the envelope, while Cryptographic Algorithm B for signing credentials. However, they both belong to the cryptographic primitives as cross-layer considerations. This subsection discusses the cross-layer considerations, which also finalizes the reference model. Figure \ref{fig:layers} depicts the complete view of the reference model. 

\subsubsection{Authentication and Authorization}
Authentication and authorization enable users to access resources and services in the digital world. Authentication of a user is the process of verifying a claimed identity, whereas authorization is granting access privileges \cite{stouffer2015guide}. These identity management processes evolved into dedicated identity management systems (IDMS). Depending on the use case and vertical, it may be necessary to build transitionary SSI solutions, which bridge the gap between the existing IDMS and the SSI paradigm\cite{Yild2109:Connecting}.

Originally, IDMS was based on centralized infrastructures where an organization managed and controlled authentication and authorization. However, centralized authentication and authorization management have the disadvantage that a user must create a separate identity for each organization and control the user data. Further development of centralized IDMS is called federated IDM systems. Federated IDMS provides users with single sign-on to multiple distinct organizations without storing identity data at each organization. However, in centralized and federated IDMS, user-related identity data is stored and controlled by the residing organizations \cite{9272223}. In contrast to centralized and federated IDMS, SSI enables users to authenticate and authorize online services without storing identity information in the IDMS of organizations.

Authentication in an SSI technology stack occurs in every layer. On the technical trust layer, authentication is necessary to execute CRUD operations on a DID document by using the private key of the public/private key pair mentioned in the verification property of the DID document. On the agent layer, agents exchange DIDs, which then can resolve into DID documents. Agents take the public key mentioned in the keyAgreement property to communicate securely with each other. The sender encrypts the message with the private key of the pubic/private key pair (authentication) mentioned in the keyAgreement of their DID document, which is cryptographically verifiable by the receiver. Issuers and holders authenticate on the credential layer by signing VCs and VPs, respectively. On the application layer,  applications such as Trust Ping rely on message exchange with a message envelope. Therefore this exchange requires the same authentication required on the agent layer. Optionally, authentication might be necessary in specific verticals and use cases. 

In an SSI technology stack, authorization occurs in the credential layer. For example, via VPs, identity holders assert claims to verifiers so they can be validated to grant access to goods and services. The access the holders might be granted depends on the authorization based on the asserted claims. Optionally, authorization might occur on the application layer depending on the use case and vertical.

\subsubsection{Attachments}
Some technology stacks, such as the Hyperledger Aries, support attachments for exchanging DID documents and credentials. From its early age, the Aries framework supported attachments by enabling attachments to DIDComm messages using one of the following three methods\cite{aries17attachments}:

\begin{itemize}
    \item \textit{Inlining}: DIDComm attachments are inlined when data is assigned as a value and a JSON key. For example, a message about arranging a rendezvous may include information about the location. One of the supported formats for inlined data is the Google Maps pin format. The inlined attachment is also a structure with meaning at rest, separate from the message it conveys, and the versioning of the structure of that message may not depend on the versioning of the protocol used to meet. Inlined data must be in JSON format since any other format would break the rules.

    \item \textit{Embedding}: Embedding an attachment is a method that allows any data to be attached with an attachment descriptor. By convention, this type of field ends with \texttt{\textasciitilde attach} suffix, which adds semantics to the field itself by using it as a field decorator\cite{aries11decorators}. Embedded attachments are not human-readable and are usually encoded with base64url. Perhaps a considerable benefit to such an approach is that the attached data can be any MIME type instead of plain JSON, unlike inlining.

    \item \textit{Appending}: Similar to embedding, appending uses ~attach decorator. The decorator, in this case, is an array of attachment descriptor structures. Therefore, multiple attachments can be put into one decorator. 
\end{itemize}

Depending on the use case, different attachment methods are better than others. For instance, inlining is the better fit for simple and small data. For example, DID Exchange Protocol 1.0 supports inlining for DID documents in messages\cite{didexchangev1}. In contrast, embedded data is supported for issuing credentials or presenting proofs. Appended attachment types are very flexible. However, they can have problems running through semantically sophisticated processing. 

Attachments pave the road for new use-cases for DIDComm and SSI and enable various specifications and protocols to be modeled in an attachment, making the combination of attachments and Aries credential exchange protocols capable. For example, different specifications protocols such as presentation exchange or credential manifest can be integrated into the Aries Present Proof and Issue Credential protocol. In addition, attachments on DIDComm messages also enable DIDComm for P2P messaging and data exchange, which is a use case that is currently under the radar. 

\subsubsection{Disclosure Zero-Knowledge Proof}

As explained in Section \ref{sec:credential_layer}, a VC consists of one or more claims about the identity subject. Zero-knowledge proofs (ZKPs) enable selective disclosure of those claims, meaning the identity subject can disclose only a subset of the claims in a VC instead of disclosing everything in the VC. Selective disclosure is also an essential property as regulatory frameworks such as eIDAS align with some of the GDPR philosophies such as data minimisation\cite{tsakalakis2018selectivedisclosureeIDAS}. In the EU, the eIDAS 2.0 is in development to support new technologies such as VCs and ISO/IEC 18013-5:2021 mobile driver's license (mDL)\footnote{Personal identification — ISO-compliant driving licence — Part 5: Mobile driving licence (mDL) application \url{https://www.iso.org/standard/69084.html}}. Although not finalized, selective disclosure might be required for electronic identification purposes. Therefore, it is crucial to consider it, thus ZKPs while choosing the technology stack. 

\subsubsection{Storage}
In most cases, the exchanged information must be stored. For example, issued verifiable credentials must be stored in holder wallets, or DID documents of public DIDs must be stored in a VDR. This subsection is a set of considerations for storing SSI-related information in suitable mediums. Since there are many considerations and implementation options, this subsection is not exhaustive but shows some examples.

Technology stacks such as uPort store DID documents on the InterPlanetary File System (IPFS)\cite{uportintro}. In simple terms, IPFS is a peer-to-peer distributed file system that connects all computing devices through a single file system. The IPFS has some similarities to the web, but IPFS can be viewed as a swarm of BitTorrent nodes exchanging objects. IPFS provides a content-addressed block storage protocol, which allows high throughput\cite{benet2014ipfs}. uPort \cite{uportintro} offers SSI tools based on the Ethereum ledger and smart contracts running on-chain for building decentralized user-centric applications. The contracts are used to store and locate DIDs, and the DID documents are stored in IPFS\cite{uportVC}. Jolocom is pursuing its approach for implementing SSI, in which DIDs are stored as KERI self-certifying identifiers (SCID) by default, which then can be anchored to any VDR. 

The Hyperledger Indy technology stack, on the other hand, stores every transaction related to a DID in the Indy DLT\cite{indytx}. Then, these transactions and the underlying information are ordered via DID Indy method in an abstract presentation. This presentation is called a DID document. In short, Indy DLT is used for storing related information about a DID. A DID document is an abstract construct, a summary of the related information instructed by the DID method to be constructed. In addition, information regarding the network rules, governance, schema, and credential definitions are also stored in the Indy DLT\cite{indytx}.

Verifiable credentials in identity holder wallets can be stored in regular relational DBs like MySQL or MongoDB. For an Hyperledger Aries solution, the internal implementation guidelines and specification of external interfaces are defined in \cite{arieswallets}. Jolocom uses SQLite for storing verifiable credentials\cite{jolocomstorage}. In most implementations, some type of relational DB is deployed. 

\subsubsection{Compliance}
Although SSI is a decentralized approach to digital identities and the network can be distributed globally, it is still bound to the local regulations if that network is to be operated in those local jurisdictions. Therefore, design choices must be considered when creating an SSI technology stack to ensure that it complies with the regulations. For example, to comply with the GDPR in the EU, one of the requirements is that no personal data is to be processed without the data subject's consent\cite{pandit2019gconsent}. Personal information is not to be written in a public ledger as anybody can read the data stored in the ledger, and data subjects cannot consent to everybody to process their data. Therefore, the network governance must not allow storing of personal information in the ledger. A private permissioned network can be deployed as the VDR instead of a public permissioned or permissionless one. Anyhow, the regulation requires a technical design choice for compliance.

\subsubsection{Data Formats}
Data models such as VC Data Model and DID Core are written for JSON, and implementations generally support JSON or its Linked Data extension JSON-LD. However, many other data formats are officially supported for SSI technology stacks. Any data format capable of expressing the data models can be used as a data format\cite{vc-data-model}\cite{did-core}. Noteworthy mentions are XML, YAML, and CBOR. Depending on the use case, these data models might be more suitable than JSON or JSON-LD. For example, CBOR could be more suitable for IoT devices due to its smaller size compared to JSON\cite{bormann2020rfc} where the data throughput can be a bottleneck. On the other hand, if more flexibility is needed, YAML can be chosen since it is considered a superset of JSON, where JSON files are, in most cases, valid YAML files\cite{jsonvsyaml}.

\subsubsection{Timestamps}
When a verifiable credential is issued, a timestamp field describes when the issuer issues a credential to the identity holder in most implementations. However, if the timestamp is only in a verifiable credential, there is no way to ensure that the issuer issued those credentials at the time written in the timestamp field. For example, an identity holder might claim that they have owned a credential for years because it is stamped in the verifiable credential, but it might have been issued only days ago. The verifier trusts the issuer directly in this case. 

To mitigate the potential risk of fraudulent timestamps, a more trust-less approach can be implemented by timestamping hashes of documents, or in the case of verifiable credentials, in VDRs. For example, eSSIF writes the issuance dates to the unique IDs of verifiable credentials in the Revocations and Endorsement Registry\cite{ebsitimestamp}. Another way is Open Timestamps, which timestamps the hashes of verifiable credentials into the Bitcoin ledger. However, the implementation is flexible enough to support other DLTs\cite{opentimestamps}.

\subsubsection{Crypto Primitives}
\begin{figure}[!ht]
    \centering
    \includegraphics[width=.48\textwidth]{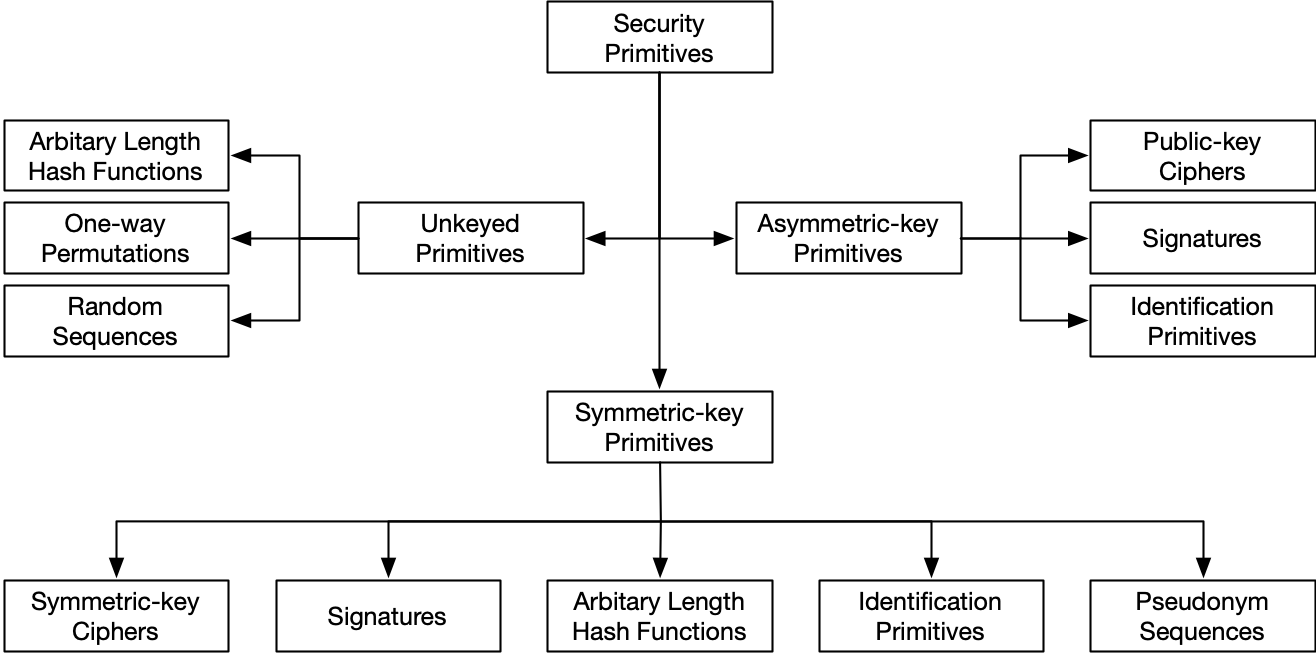}
    \caption{Composition of cryptographic primitives, taken from \cite{menezes1997} }
    \label{fig:taxonomy-crypto-primitives}
\end{figure}

Cryptographic primitives, also known as crypto primitives, are low-level cryptographic algorithms used for creating higher-level cryptographic functions\cite{cryptoprimitivenist}. As an example of this statement, JSON Web Messages can be signed with the Ed25519 digital signature scheme, which is used as asymmetric-key primitives that count as security primitives. The Handbook for Applied Cryptography \cite{menezes1997} also calls them ''basic cryptographic tools'' and presents evaluation criteria and a taxonomy, reproduced in excerpts in Figure \ref{fig:taxonomy-crypto-primitives}. 

In general, crypto primitives are implemented in crypto libraries. For instance, Hyperledger stack contains Ursa\footnote{Hyperledger URSA:  \url{https://github.com/hyperledger/ursa}} as crypto libraries, which includes, Ed25519, Secp256k1, BBS+ and CL signatures. Moreover, it is possible to include language-specific crypto libraries in a technology stack such Botan\footnote{Botan - Crypto and TLS for Modern C++: \url{https://github.com/randombit/botan}} for C++, or libsodium\footnote{libsodium:  \url{https://github.com/jedisct1/libsodium}} for JavaScript.

\section{Interoperability Considerations} \label{considerations}
The reference model can help create new technology stacks. As mentioned in Section \ref{problem}, there are multiple SSI technology stacks that serve different business needs. The reference model can also help map existing technology stacks to compare and understand the differences. As a result, organizations implementing technology stacks can align on interoperability goals (see Section \ref{taxonomy}) and understand how they can be reached with the reference model. For instance, if agents from different technology stacks must be interoperable to a certain level (i.e., semantic level interoperability), they can do so by supporting the same standards and protocols related to some of the components mentioned in the reference model. Following the structure of the reference model, we discuss the interoperability considerations to achieve those interoperability goals in this section. 

\subsection{Technical Trust Layer}
Interoperability on the technical trust layer is not a concern as long as technology stacks follow the standardizations of the components related to the layer. This is because there is a single standardization of the DID documents, methods, or protocols deployed in this layer, and no business needs to derive other standards and protocols. Therefore, the technology stacks agree on the standards of required components such as DID documents and DID methods while having divergence in optional components such as DID scaling options. The following interoperability considerations are relevant in the technical trust layer. 

\subsubsection{Alignment on DID Document Format}
To initiate technical trust between agents, DID documents must be resolvable and comprehensible by the interacting agents. For a shared comprehension of the context of a DID document, it must be standardized. The W3C standardizes the body of the DID document in the DID Core Working Group\cite{did-core}. Therefore, DID documents are not a concern of technical or syntactical interoperability as long as the technology stacks follow the W3C standardization of DID documents. 

\subsubsection{Different DID Methods}
Although there is a standardization of what a DID method must fulfill, how they are fulfilled is up to each DID method implementation. It is less of a problem when the agent supports the DID method natively, which is the case when the agent is natively a part of a technology stack such as the Hyperledger Indy. However, other agents, such as verifiers, might not be native to the technology stack but must still resolve a DID to a DID document so that it can communicate with the agent to which the DID belongs. 

There are more than a hundred DID methods and if a non-native agent wants to communicate with a DID coming from one of those DID methods, it must be able to resolve via the DID method of that DID. Native implementation of the DID method on that scale is not feasible, and so is adding new DID methods. As a solution, DIF universal resolver has been developed to offer a gateway to resolve supported DID methods\cite{uniresolver}. A universal resolver driver needs to be implemented in the DID method, and the DID method itself is added to the universal resolver repository. Resolvability of DID methods is crucial for achieving technical interoperability.

\subsubsection{Different DID Scaling Options}
DID scaling options such as KERI and Sidetree and not natively interoperable. Therefore, if the technology stacks require DID scaling options and strive to achieve technical interoperability, they must support each other's DID scaling options. For instance, if one stack is scaling their DID operations with KERI and the other with Sidetree, they must support both solutions to execute CRUD operations. 

\begin{figure*}[t]
    \centering
    \includegraphics[width=\textwidth]{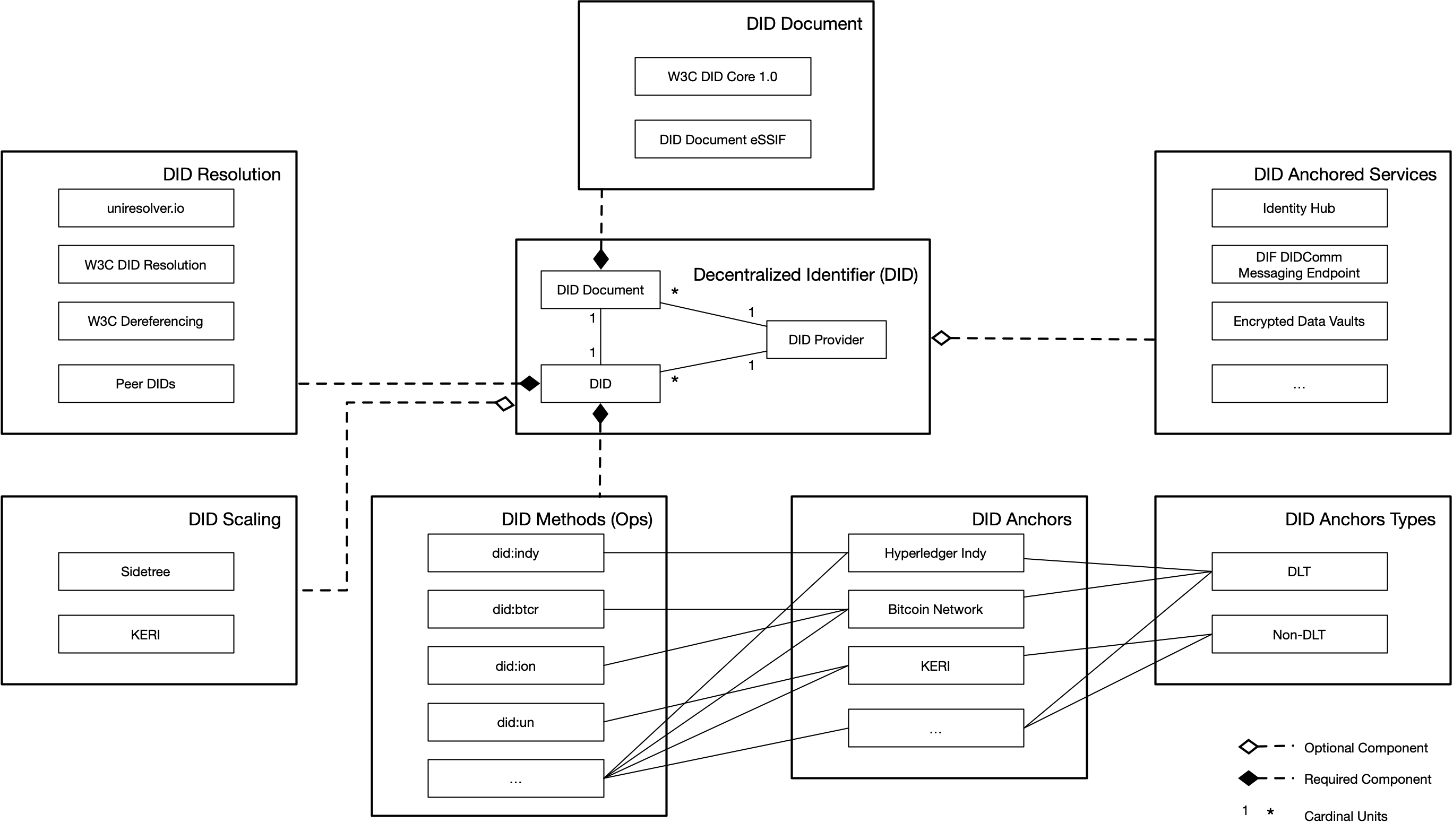}
    \caption{Public Trust Layer Composition}
    \label{fig:DIDTax}
\end{figure*}

Interoperability considerations and the components on the technical trust layer of the reference model provide complete information related to the creation of the technical trust. Figure \ref{fig:DIDTax} depicts the connections between the components and considerations in the technical trust layer with the DID as the point of reference.

\subsection{Agent Layer} \label{agent_Interop_Considerations}
Depending on the technology stack and business needs, there are divergences in several components of the agent layer. Therefore, interoperability on the agent layer becomes more challenging due to the necessary alignment between different technology stacks. The technology stacks willing to be interoperable must either choose the same variation of components or support the variation of each other. In the following, we discuss the interoperability considerations on the agent layer. 

\subsubsection{Different Message Envelopes}
In addition to the two versions of the DIDComm message envelope we discussed in the reference model, another envelope is a part of the OIDC technology stack. This contender for the message envelope is called the Self-Issued OpenID Provider (SIOP) v2. It extends the OIDC stack by allowing DID keys and discovery metadata to create a secure envelope between agents while fulfilling the information security properties mentioned above. The public key is shared as a JSON Web Key, and the DID method for resolving key material is supported as an extension to the OIDC. The key material can be exchanged using the native discovery services of OIDC or with out-of-band mechanisms\cite{siopv2}.

Moreover, there are two versions of DIDComm. DIDComm was initially created by the Hyperledger Aries project to facilitate agent-to-agent communication in a decentralized fashion \cite{aries2019didcomm}.
The technology was tailored to the Aries implementation's specifics and Hyperledger Indy, which contains the components libraries related to the technical trust and credential layers. Thus it lacked the flexibility for a general application in other scenarios. Therefore, in 2019, the DIF decided to work on the DIDComm standardization as an open standards approach outside the Hyperledger technology stack. As the specification brought forward by the DIF \cite{hardman2019didcomm} introduces certain incompatibilities and breaking changes\cite{didcommv2whatsnew} compared to the Indy version, the two specifications are referred to as \textit{DIDComm v1} and \textit{DIDComm v2}. The differences between the two versions are, among other things, the handling of DIDs, underlying crypto primitives, and message structures.

The choice of message envelope has a substantial impact on technical interoperability. For example, the message structure between the DIDComm v1 and v2 versions varies to the point that they are incompatible with one another\cite{didcommBasicMessage}. Furthermore, SIOP v2 comes with a different message structure as well. As a result, none of the message envelopes are compatible with each other. Therefore, the same message envelope must be supported to exchange information between agents. If two technology stacks utilize different envelopes, they must support each other's envelope to achieve technical interoperability on the agent layer. 

\subsubsection{Different Transports}
Similar to the envelope, transport is another vital component for achieving technical interoperability. When exchanging information, the agents must support the same transport protocols. For instance, if Agent A wants to communicate with Agent B, which only supports HTTP, then Agent A can only send a message to Agent B via HTTP.

\subsubsection{Support of Same Control Recovery Methods for Portability}
Control recovery does not interfere with any level of interoperability regarding the interactions between SSI agents. It only affects the SSI actors if they want to change the agent and migrate their keys to a new one. However, since the BIP-39 is the industry standard and is widely adopted, control recovery is not an interoperability concern for key migration and portability as long as the new agent supports BIP-39.  

\subsubsection{Key Operations}
How the keys are operated might affect legal interoperability between the SSI actors. For instance, a verifier might require a high level of assurance that requires the key operations to be isolated from the users' direct access. The verifier might deny access to its services if this requirement is not fulfilled. There are multiple ways to isolate the keys from the user. To have a clear understanding of different options, we discuss various methods of key operations, which all come with a trade-off between security and convenience: 

\paragraph{User has direct access to keys} 
In its simplest form, users have direct access to their keys and key recovery methods like the BIP-39 and other DKMS methods. The agent has direct access to the keys and signs transactions without complications. This approach comes with convenience but with some open security issues. First, users can share their keys with other people, using the same keys and digital identities to access goods and services. Second, they are susceptible to theft which can be impactful. Moreover, some use cases, such as official eID solutions, only allow a certain level of assurance. The higher the level, the less it is possible to have direct access to the keys. 

\paragraph{Keys are isolated from the user}
The keys can be isolated from the user while still being stored in the user's domain to solve the security issues mentioned above and reach a higher level of assurance. The keys are accessed through an interface only to get the value of calculations or signatures from the isolated system. Key isolation is not a new concept but can be found in other telecommunication components, such as SIM cards, where the user possesses the card but does not have access to how it communicates with the home network. There are multiple options to isolate the keys from the user's direct access.

\textit{Hardware Secure Module (HSM):}
An HSM is an isolated component of a hardware device independent of other systems, such as a smartphone. It is isolated from the code within the other systems and is resilient to tampering, theft, or manipulating key operations. They can encrypt and decrypt digital signatures (or messages) without the private keys being exposed out of the module. HSMs have processors that are capable of specific cryptographic algorithms and curvatures. Due to lack of demand, algorithms capable of ZKPs are known to be not supported by HSMs. Nevertheless, there are options to execute code in HSM that is not native to the module. One such way is using a JavaCard-Applet, where unsupported signatures can be executed in any HSM. 

\textit{Cloud HSM:}
A Cloud HSM is a security module that runs on a cloud service such as AWS or Azure. Users can encrypt or decrypt data using various algorithms\cite{cloudHSMhuang2018survey}. In the case of SSI, a cloud HSM can be instantiated for cloud agents. 

\textit{Trusted Execution Environment (TEE):}
TEE is an isolated secure area of a device's central processor. Intel's Software Guard Extensions (SGX)\footnote{Intel Software Guard Extensions: \url{https://www.intel.com/content/www/us/en/developer/tools/software-guard-extensions/overview.html}} and ARM TrustZone\footnote{Trustzone for Cortex-A: \url{https://www.arm.com/technologies/trustzone-for-cortex-a}} are two examples of TEE. They guarantee specific characteristics of information security, including integrity, authentication, and confidentiality. It is an isolated environment that offers a higher level of trust for the executed applications. Since TEEs are a part of the central processor, they are flexible with the code that can run on them. Unlike HSMs, TEEs do not run code related to cryptography exclusively. 

\textit{Trusted Platform Module (TPM):}
According to the Trusted Computing Group's (TCG) specifications, a TPM is a chip that adds basic security features to a computer or similar device\cite{tpm}. For example, TPM can be used on a PC to secure cryptographic keys, digital rights management, and Windows Domain logon.

\textit{NFC and Smart Cards:}
The final type of isolating the keys from the user can be found in the current implementation of the German eID solution. The AusweisApp2\footnote{AusweisApp 2: \url{https://www.ausweisapp.bund.de/en/ausweisapp2-home/}} allows users to see the data stored in their eID and send them through a secure channel for verification by reading information stored within the ID chip and communicating a subset of this information via NFC. Users own both the smartphone and the physical ID card, but the information on the physical ID card can not be manipulated from the outside.

As mentioned earlier, key operations play a role in terms of legal interoperability between SSI agents. However, the level of assurance can be substituted with compatibility as the code can run in a normal environment without isolation if the signature or encryption can not be processed by the HSM or TEE, albeit with a lower level of assurance. It is also worth mentioning that interoperability of HSMs with some of the signature algorithms, such as algorithms capable of ZKPs, is de facto an unsolved problem. Therefore, achieving legal interoperability with a verifier requiring a higher level of assurance for credentials with ZKPs without giving away privacy preservation is currently not possible.

\begin{figure*}[t]
    \centering
    \includegraphics[width=\textwidth]{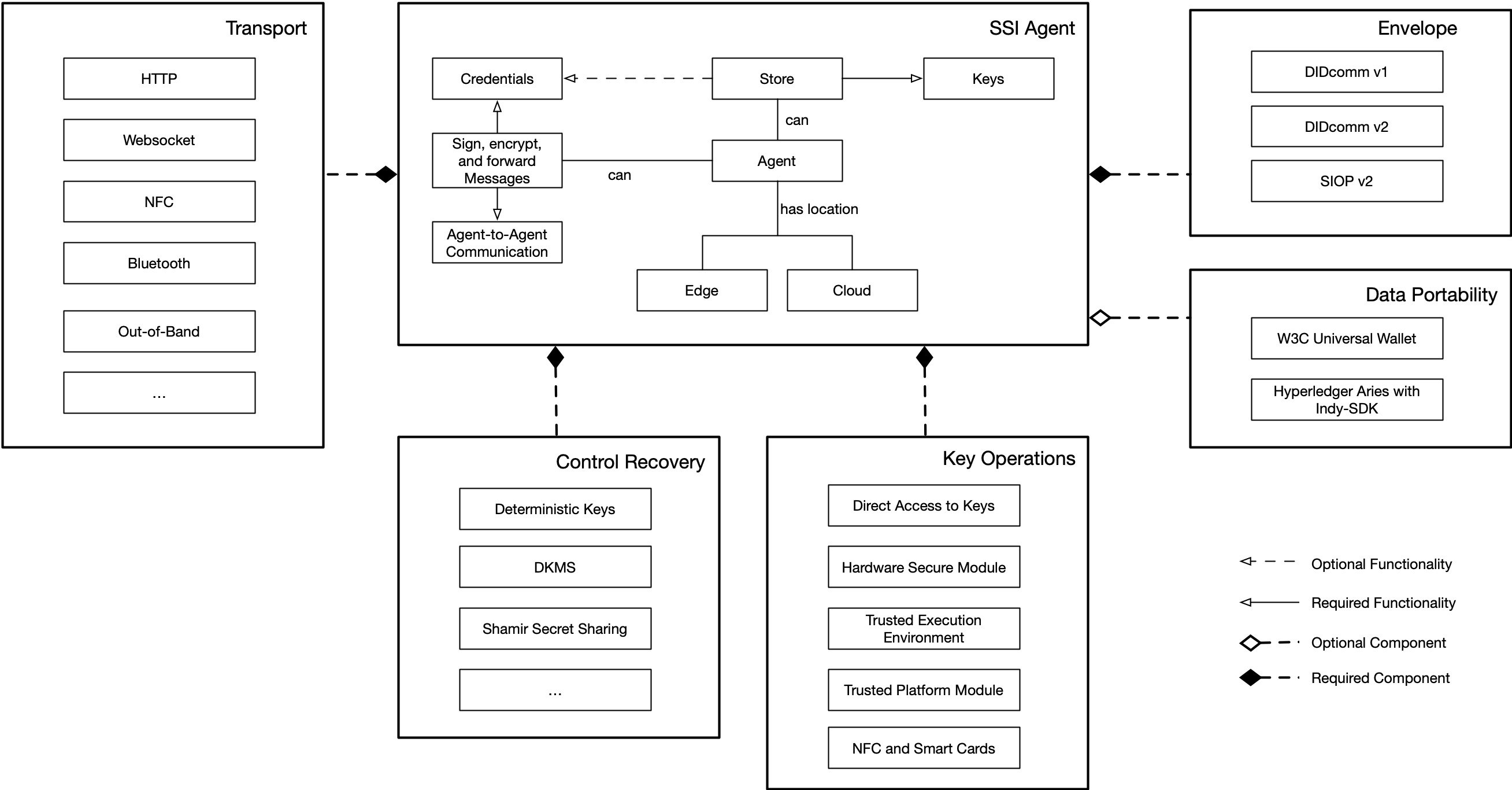}
    \caption{Agent Layer Composition}
    \label{fig:AgentTax}
\end{figure*}

Interoperability considerations and the components on the agent layer of the reference model provide complete information on this layer. Figure \ref{fig:AgentTax} summarizes the composition of the agent layer with the agent as the point of reference. 

\subsection{Credential Layer}
Due to all components having more than one standardization and protocol, the credential layer proves to have the most complex interoperability problems between technology stacks as they include one set of standards and protocols for each component. For instance, the Technology Stack A supports Proof Type A, Revocation A, Credential Exchange A, and Credential Binding A, while Technology Stack B supports the B variants of the mentioned components. In that case, as long as they do not support each other's standards and protocols, they will not be interoperable. Moreover, even the slightest deviation from the technology stack causes interoperability problems. For example, if Technology Stack B supports Proof Type A, Revocation A, and Credential Binding A but still only supports Credential Exchange B, the Technology Stack A and B will still be not interoperable. 

In the following, we discuss the divergences in standards and protocols on the components of the credential layer and explain why interoperability becomes a problem with diverging protocols and standards. 

\subsubsection{Different Credential Formats}
We discussed the most adopted credential formats in the reference model according to the W3C data model. However, several other credential formats are relevant and found in some technology stacks. To achieve technical interoperability, agents exchanging credentials must support the same credential formats.

As a variation of verifiable credentials, anonymous credentials (\textit{AnonCreds}) are the implemented credential format in the Hyperledger Indy technology stack. AnonCreds do not follow the standardization of verifiable credentials according to the W3C. Due to interactions that must be modeled in the specification and the required data structure being divergent from the W3C standard, AnonCreds are currently standardized and in draft status. AnonCreds are heavily privacy-focused and are the only implementation of credential proof types that enable predicates\cite{anonCreds}.

Moreover, as a part of the OIDC technology stack, \textit{OIDC ID token} is sent to a relying party after user authentication\cite{oidcidtoken}. It is encrypted using JWT consisting of a header, a payload, and a signature. The header contains metadata such as encryption method and algorithm. The payload contains information regarding the identity holder. Similar to the verifiable credentials, the payload is tamper-resistant, and its origins are cryptographically verifiable, which can be verified by the signature\cite{jwt}. OIDC mainly uses it for authentication purposes in the centralized and federated identity access management systems and decentralized identities. When used with decentralized identities, aa \textit{OIDC VP token} is passed along with the \textit{ID token}, which contains the representation of a verifiable presentation in JSON format\cite{oidc4vp}. 

Lastly, \textit{Open Badges} are image files that contain information about digital learning achievements. Similar to verifiable credentials, they are tamper-resistant, and the issuer is cryptographically verifiable. They are based on a set of specifications and open standards\cite{openbadgesreport}. When an open badge is issued to a subject, a portable image containing verifiable metadata related to the issuer is created. This image can be displayed by any open badge compatible application\cite{openbadgeswhat}.

\subsubsection{Different Proof Types}
In addition to different credential formats, the interacting agents must support the same proof types to issue, store, and validate self-sovereign identities. Similar to the credential formats, there is more than one proof type. The divergence of proof types is a result of different business requirements. For instance, a proof type might be more privacy-preserving but complex, while another supports attachment of vocabularies but is less privacy-preserving. In the following, we discuss the four predominant proof types found in VCs, VPs, and AnonCreds. 

Although a VC or VP can be modeled in data formats such as CBOR and XML, JSON and JSON-LD are the most used data formats. JSON stands for JavaScript Object Notation and is one of the most prominent data interchange formats for exchanging data between platforms and systems. Due to its language-agnostic nature, many systems and architectures agree upon using JSON as a standard to exchange data to simplify the data exchange between those systems and architectures \cite{JSONbassett2015introduction}. In the case of SSI, self-sovereign identities are contained in JSON(-LD) to be exchanged between agents. Two of the four flavors use JSON as the data exchange format.

The first type of credential proof is JWS in JSON Web Tokens (JWT), an open standard for securely exchanging information between parties in JSON format, which can be digitally signed with a private key of a public/private key pair. Similar to the OIDC ID token mentioned earlier, a JWT contains a header, a payload, and a signature\cite{JWTsheffer2020json}. In the case of VCs, the private key of a public/private key pair signs the payload (metadata and claims) as proof of origin in a tamper-resistant manner. In the case of VPs, the private key of a public/private key pair related to the identity holder DID signs the payload (VC). Both JSON and JWT are commonly found on the web, and web authentication, respectively, and they count as the most straightforward flavor among the four \cite{young2021}. Listing \ref{listing:jwtexample} is an example of a VC encoded as a JSON Web Token, which can be decoded via services such as \url{https://jwt.io/}. 

Depending on the specification, different signature algorithms and specifics can be supported. For instance, while the Internet Assigned Numbers Authority (IANA) supports a wide variety of algorithms\cite{iana}, the Internet Engineering Task Force (IETF) defines a smaller subset as standard in the RFC-7518\cite{jwa}. Even a smaller set of algorithms are supported in the JWS Signature Suite according to the W3C\cite{jwsc2c}. Therefore, the choice of signature standards and suites plays a crucial role in achieving technical interoperability.

\begin{lstlisting}[caption=A VC as JWT that can be decoded via a JWT decoder, label=listing:jwtexample, captionpos=b]
eyJ0eXAiOiJKV1QiLCJhbGciOiJFUzI1NksifQ.eyJzdWIiOiJkaWQ6Z
XhhbXBsZTpzdHVkZW50RElEIiwibmJmIjoxNjM2OTkyNDIyLCJpc3MiO
iJkaWQ6ZXhhbXBsZTppc3N1ZXJESUQiLCJ2YyI6eyJAY29udGV4dCI6W
yJodHRwczovL3d3dy53My5vcmcvMjAxOC9jcmVkZW50aWFscy92MSIsI
mh0dHBzOi8vdzNpZC5vcmcvc2VjdXJpdHkvc3VpdGVzL2p3cy0yMDIwL
3YxIl0sInR5cGUiOlsiVmVyaWZpYWJsZUNyZWRlbnRpYWwiLCJCYWNoZ
WxvckRlZ3JlZSJdLCJjcmVkZW50aWFsU3ViamVjdCI6eyJmYW1pbHlOY
W1lIjoiRG9lIiwiZ2l2ZW5OYW1lIjoiSmFuZSIsImFsdW1uaU9mIjoiV
FUgQmVybGluIiwiYXdhcmQiOiJCYWNoZWxvciBvZiBTY2llbmNlIn19f
Q.RbAPINtFFDujOOW1m8PwKc5AgD8Zel-I0-8fC_pJ8mhhH5-a8HLDmc
R8VZu_Hy1rRfrssTETiypPUPPWB0JPDQ
\end{lstlisting}

Although JWT is widely used and simple, it lacks some properties which might be required for specific use cases. For example, while obtaining the verifiable credential, it does not have privacy-preserving features and gives away a persistent identifier, a DID of the credential subject. Furthermore, these credentials also disclose either everything or nothing, meaning in most cases, the identity holder shows more information to the verifier than necessary. Zero-knowledge proofs (ZKP) can help eliminate these two shortcomings. One type of credential proof is created with Camenisch-Lysyanskaya (CL) Signatures, which enables ZKP\cite{young2021}. 

Proposed in 2002, CL Signatures allow selective disclosure, meaning the credentials include only selected name-value pairs instead of disclosing everything or nothing. There is also no need to reveal a persistent identifier to prove the ownership of the credentials. Instead of showing a full signature and a persistent identifier, a proof of ownership is revealed as a proof value without revealing the exact value (link secret) of the proof of ownership. This property also enables the option to derive a VP from multiple VCs as they are all bound to the same secret. Jan Camenisch and Anna Lysyanskaya describe this type of credentials as Anonymous Credentials \cite{camenisch2002signature}, which gave the name to the credentials (AnonCreds for short) instantiated in the Hyperledger Indy technology stack. Besides these two characteristics, CL Signatures are capable of predicates, enabling credential subjects to prove that they at least 18 year old without giving away how old they are. Another characteristic of this proof type is having a structural disambiguity, meaning the issued and presented credentials come from a credential definition, which might be beneficial for some use cases.

\begin{lstlisting}[caption=An Anonymous Credential with JSON format and CL Signature, label=listing:jsoncl,captionpos=b] 
{
  "schema_id": "did:example:issuerDID:2:degree schema:23.61.16",
  "cred_def_id": "did:example:issuerDID:3:CL:115612:TUB.agent.degree_schema",
  "key_correctness_proof": {
    "c": "1103503870767469451131116849332707621348740
    44888695878146031536959590209770914",
    "xz_cap": "19436439432181908959457977800460981765
    461625019228265073101245.....15218401523893378299
    01638439622695703136361232926271309117643219879",
    "xr_cap": [
      [
        "degree",
    "341008229222691682559089928533019481708336450805
    384460560355414536020.....62547041468871664367553
    998459359448173998265119212913363752614380080697"
      ],
      [
        "name",
    "196272939796096557204699175446233092755860829304
    5489600808784536861.....7379265485645096326784331
    713082650018238239026462295034203019699304937778"
      ],
      .....
      
    [
        "master_secret",
    "222519953903207933803102844436799001441319989981
    07975799402999480.....165954024457790711508903629
    464922182481493454888148338859329613903331541995"
      ],
    [
        "date",
    "293321842670898730162110771120029104264500535338
    6361139801805446250.....3410841150862859981113603
    618246179556767024644417997315095113970401156872"
      ],
      [
        "timestamp",
    "990003338110938061501902038064480651067904311425
    624227076485850490.....06189644124378266561105970
    683285429329859345646233455026580235422045454372"
      ]
    ]
  },
  "nonce": "151499423063318703466275"
}
    
\end{lstlisting}

JSON is quite efficient in serializing and processing data on its own. However, the capabilities of semantic disambiguity and linking data outside of the document are missing in plain JSON. JSON and Linked Data have been merged into JSON-LD to add these capabilities. JSON-LD is a data exchange format and an extension to JSON for serializing Linked Data in JSON format that allows the JSON to be interpreted as Linked Data. It is intended to use Linked Data in web-based environments and build context-aware interoperable web services\cite{json-ld}. In the case of SSI and credential formats, it enables the storage of context in a web server or similar, which allows the mapping of key-value pairs to an RDF ontology and creation of semantic disambiguity\cite{young2021}. The last two of the four flavors are in JSON-LD data exchange format. 

The type of proofs created for JSON-LD data exchange format is called data integrity proofs, which are deployed to verify the authenticity and integrity of Linked Data documents. In the case of VCs and VPs in JSON-LD format, they prove the origins or the ownership of the credentials or presentations, respectively\cite{LDProofsW3C}. It is a powerful addition to regular JSON-type proofs because it enables open-world data modeling and semantic disambiguation. According to the Verifiable Credential Implementation Guideline from W3C, there are advantages of using data integrity proofs instead of the other proof types mentioned above, including signature chaining, canonicalization requiring only base-64 encoding, and the ability to express the proofs in other data formats such as YAML and CBOR\cite{VCguidelines}. According to the W3C specification, this type of proof must follow the \textit{JsonWebSignature2020} signature suite\cite{jwsc2c} or any of the specified cryptographic primitives.

\begin{lstlisting}[caption=A verifiable credential with JSON-LD format and LD Signature, label=listing:jsonldvc,captionpos=b] 
"@context:" [
  "http://schema.org"
],
"id": "http://example.edu/credentials/bachelor",
"@type": "Person",
"issuer": "did:example:issuerDID",
"issuanceDate": "2021-11-23T18:00:00Z",
"credentialSubject": {
  "id": "did:example:studentDID",
  "familyName": "Doe",
  "givenName": "Jane",
  "alumniOf": "TU Berlin",
  "award": "Bachelor of Science"
},
"proof": {
  "type": "JsonWebSignature2020",
  "created": "2021-11-23T18:00:00Z",
  "proofPurpose": "assertionMethod",
  "verificationMethod": "did:example:issuerDID#key-1",
  "jws": "eyJhbGciOiJFZERTQSIsImI2NC
  I6ZmFsc2UsImNyaXQiOlsiYjY0Il19..
  Cky6Edasq24as"
}
\end{lstlisting}

So far, we have discussed JSON and JSON-LD data exchange formats and suitable proof types for these data exchange formats, including a simple or a privacy-preserving proof for JSON and a non-privacy preserving proof for JSON-LD. The final flavor is another data integrity proof that comes with privacy-preserving features while having the benefits of Linked Data and semantic disambiguity. This type of data integrity proof consists of BBS+ Signatures, specifically BBS+ Signature Suite 2020, a set of parameters that enables data integrity proof conformity according to the W3C draft. Among them are BLS12-381 curvature, Blake2b statement digest algorithm, and URDNA2015 canonicalization algorithm\cite{BBS+}. Furthermore, BBS+ Signatures can create ZKPs, therefore supporting privacy-preserving features such as selective disclosure and the non-necessity of revealing a persistent identifier with a specific signature suite\cite{BBS+Matter}. Because of this, they also count as ZKPs that fulfill the requirements of data integrity proofs. However, BBS+ Signatures are not capable of predicates, which might be required for specific use cases.

\begin{lstlisting}[caption=A VC with JSON-LD format
and BBS+ Signature, label=listing:jsonldbbs+,captionpos=b] 
"@context:" [
"http://schema.org"
],
"@type": "Person",
"familyName": "Doe",
"givenName": "Jane",
"alumniOf": "TU Berlin",
"award": "Bachelor of Science"
},
"proof": {
  "type": "BbsBlsBoundSignature2020",
  "created": "2021-11-23T18:00:00Z",
  "proofPurpose": "assertionMethod",
  "verificationMethod": "did:example:issuerDID#key-1",
  "proofValue": "12315qwfdg3q352tsg4wgse6&sefs43qqrf3sw+3qa2ea3r"
  "requiredRevealStatements": [ 3, 4 ]
}
\end{lstlisting}

As these proof types show, no one best proof is suited for everything. Therefore, it is essential to choose the proof type by business requirements. Table \ref{tab:proofs} depicts an evaluation of these proof types by their properties. For example, if the use case requires ZKP, the CL and BBS+ signatures are the only contestants. On the other hand, CL signatures (AnonCreds) are currently the only implementation if predicates are required.

\begin{table*}[]
\caption{Overview of Credential Proof Types and Signatures, modified from \cite{young2021info}}
\label{tab:proofs}
\resizebox{\textwidth}{!}{%
\begin{tabular}{@{}|l|l|l|l|l|l|@{}}
\toprule
\textbf{Characteristics} & \textbf{JSON JWT} & \textbf{\begin{tabular}[c]{@{}l@{}}JSON CL Signatures\\ (AnonCreds with Indy)\end{tabular}} & \textbf{JSON-LD BBS+} & \textbf{JSON-LD JWS} & \textbf{Remarks} \\ \midrule
Semantic Disambiguity & No & No & Yes* & Yes* & \begin{tabular}[c]{@{}l@{}}*= It's possible when the ontology\\ is connected via @context\end{tabular} \\ \midrule
\begin{tabular}[c]{@{}l@{}}Structural Disambiguity /\\ Cred. Def. Necessary?\end{tabular} & No & Yes & Yes & Yes &  \\ \midrule
\begin{tabular}[c]{@{}l@{}}Usability, when die source\\ of Cred. Def. is offline?\end{tabular} & Yes* & No** & No** & No** & \begin{tabular}[c]{@{}l@{}}*= No Cred. Def. necessary\\ **= Not possible without caching\end{tabular} \\ \midrule
Predicates & No & Yes & Yes* & No & \begin{tabular}[c]{@{}l@{}}*= Not yet implemented, with\\ additional work possible\end{tabular} \\ \midrule
Selective Disclosure & No* & Yes & Yes & No* & \begin{tabular}[c]{@{}l@{}}*= It is possible using merkle trees\\ but currently no implementation\end{tabular} \\ \midrule
Credential Binding & Yes (DID) & Yes (Link Secret) & Yes (DID/Link Secret*) & Yes (DID) & *= With BbsBlsBoundSignature2020 \\ \midrule
\begin{tabular}[c]{@{}l@{}}Deriving a VP from\\ Multiple VCs\end{tabular} & Yes* & Yes & Yes* / Near Future** & Yes* & \begin{tabular}[c]{@{}l@{}}*= If the VCs have the same DID in\\ the credential subject (not recommended)\\ **= With BbsBlsBoundSignature2020\end{tabular} \\ \midrule
\begin{tabular}[c]{@{}l@{}}Standardized \\ Signature Suites\end{tabular} & \begin{tabular}[c]{@{}l@{}}IETF RFC-7518\\ JOSE ...\end{tabular} & \begin{tabular}[c]{@{}l@{}}CLSignature2019\\ AnonCredDerivedCredentialv1\end{tabular} & \begin{tabular}[c]{@{}l@{}}BbsBlsSignature2020\\ BbsBlsSignatureProof2020\\ BbsBlsBoundSignature2020*\end{tabular} & JsonWebSignature2020 & \begin{tabular}[c]{@{}l@{}}*= Credential binding without\\ persistent identifier (DID)\end{tabular} \\ \midrule
Algorithms and Parameters & \begin{tabular}[c]{@{}l@{}}HS256, HS512,\\ ES512, ...\end{tabular} & Strong RSA assumption & \begin{tabular}[c]{@{}l@{}}Pairing-friendly Curves, such as\\ BLS12-381\end{tabular} & \begin{tabular}[c]{@{}l@{}}Ed25519, secp256k1 \\ RSA, ...\end{tabular} &  \\ \bottomrule
\end{tabular}%
}
\end{table*}

The SSI community discusses finding the one proof type that will unify the solutions for years, which might not happen due to the diversity of business requirements. For achieving technical interoperability, it is vital to either align on the proof type or support a wide array of proof types. In both cases, it is crucial to implement them with defined and standardized signature suites.  

\subsubsection{Different Revocation Methods}
Another hindrance to achieving technical interoperability is different revocation methods. There are several approaches to credential revocation for VCs. Some have more privacy-preserving features than others, and some have better scalability than the privacy-preserving revocation mechanisms. DIF Revocation working group is collecting these revocation mechanisms and aiming to create a couple of standards to enable interoperability on credential revocation\cite{revocationWG}. Like proof types, agents must support the same revocation mechanisms to achieve technical interoperability. The following examines the most prominent revocation mechanisms in SSI technology stacks.

The Verifiable Credential Status Revocation List is a lightweight and straightforward specification for a revocation mechanism. It utilizes a bitstring to store a bit in a position in the string. The VC and VP contain the information regarding the location of the related revocation bit in the bitstring. If it is a 1, it means the credential is revoked. If it is a 0, the credential is not revoked. A 16KB bitstring can contain uncompressed 131,072 entries, and therefore it is not only a fast approach to validate revocation status but also a highly scalable one\cite{vc-status-rl}. The downside of this approach is the lack of privacy, as the VP needs to contain an identifier that points to the location in the bitstring, leaving personally identifiable information behind with each shown presentation. 

In the Hyperledger Indy technology stack, credential revocation has been solved while privacy concerns of the revocation mechanisms such as the Verifiable Credential Status Revocation List are eliminated. The AnonCreds v1 is a credential format with a combination of verifiable credentials and an optional revocation mechanism. Based on the paper by Jan Camenisch et al. proposing an efficient revocation for Anonymous Credentials\cite{camenischrevocation}, credential revocation in Hyperledger Indy contains cryptographic accumulators and tails files to prove the revocation status without the necessity of revealing an identifier\cite{indy-revo}. 

An accumulator can be thought of as a product of multiple prime numbers. For example, assuming that \texttt{a * b * c * d = e}, \texttt{e} would be the accumulator. Since \texttt{a}, \texttt{b}, \texttt{c}, and \texttt{d} are prime numbers, if any of those numbers are removed, the accumulator can't be divided into the removed prime number without getting a decimal number. A tails file is a list of prime numbers multiplied by each other to create the accumulator. Revoked entries in tails files are not multiplied to get the accumulator\cite{indy-revo}.

A revocation-enabled credential has a private factor, one of the prime numbers in the tails files. The rest of the prime numbers to calculate the accumulator are called a witness. The prover must demonstrate that her credential has not been revoked when proving her credentials. She provides the mathematics that shows the accumulator value from the witness plus her factor. She does this in a way that does not divulge her private value to avoid correlation\cite{indy-revo}. This type of revocation is the industry standard of privacy-preserving revocation mechanisms. However, the downside is that the tails files can be as large as 1 GB while allowing roughly 100.000 credentials to be in the tails file. Therefore, this revocation mechanism is highly privacy-preserving but not scalable.

Various scalable and privacy-preserving revocation options are in development at the DIF Applied Cryptography Working Group\footnote{DIF Applied Crypto Working Group:  \url{https://identity.foundation/working-groups/crypto.html}}. A noteworthy option is Zero-Knowledge Signed Accumulator Memberships (zk-SAM), which uses a similar approach to AnonCreds revocation while being more scalable, and compatible with BBS+ Signatures\cite{zk-sam}. Another consideration is a privacy-preserving revocation proposal, which proposes using two cryptographic accumulators instead of one found in AnonCreds\cite{lodder-revo}. There are other considerations, but all of them, including the noteworthy mentions above, are still in the early development and have not seen scaled implementation.

\subsubsection{Different Credential Exchange Protocols}
The exchange of credentials follows specific steps and states for credential issuance (VC) and proof presentation (VP), and there is more than one protocol that enables credential exchange between agents. Like the first three components in the credential layer, agents must support the same exchange protocols to achieve technical interoperability. In the following, we discuss the different credential exchange protocols.

Aries Issue Credential Protocol 2.0 (Aries RFC 0453) is a protocol that formalizes the messages used during the issuance of credentials independent of the credential proof types\cite{issuecred2}. This issue credential protocol divides the procedure of credential issuance into sub-categories and consists of the following messages:

\begin{itemize}
\item \texttt{propose-credential}:
    Identity holders tell the issuer which claims they wish to receive. Self-attestation to a new credential can be proposed through this message. 
\item \texttt{offer-credential}:
	This message sends the credential offer containing what is to be issued. It is an optional requirement for some implementations. 
\item \texttt{request-credential}:
    If none of the two messages above are sent, request-credential is the message that starts the protocol.
\item \texttt{issue-credential}: is the actual payload that delivers the verifiable credential to the identity holder.  
\end{itemize}

The protocol is deliberately designed not to involve any particular proof type or specifics. The issued credential format or credential proof is delivered in attachment format\cite{issuecred2}. Aries RFC Attachments describes the supported attachment formats for the Aries Issue Credential Protocol 2.0\cite{attachments}.

In version 1.0, the Aries Issue Credential protocol was mainly focused on AnonCreds\cite{issueCredv1}, and the Aries Issue Credential 2.0 is a significant step towards offering an open protocol for any credential format and proof type while still depending on the DIDComm v1 message envelope. Issue Credential 3.0 follows the same openness towards different proof types while supporting DIDComm v2 message envelope\cite{issuecredv3}. 

Another type of credential exchange is for sending and verifying verifiable presentations. The Hyperledger Aries offers another protocol similar to Issue Credential Protocol 2.0 by taking an agnostic approach for proofs and credential formats. The protocol is called Aries Present Proof Protocol 2.0 (Aries RFC 0454). Unlike the first version of the protocol, which was focused on AnonCreds\cite{presentproofv1}, the Present Proof Protocol 2.0 supports any DIDComm v1 attachments\cite{presentproof2}, thus supporting various proof types. In addition, Present Proof Protocol 3.0 supports different proof types with DIDComm v2 message envelope\cite{presentproofv3}. For the interactions between the identity holder and the verifier, three message types are necessary:

\begin{itemize}
\item \texttt{propose-presentation}:
    An optional message for the identity holder to propose a presentation or to send a counter-proposal as a response to the request-presentation message from the verifier
\item \texttt{request-presentation}:
	A verifier requests a presentation from an identity holder. 
\item \texttt{presentation}:
    The identity holder provides a presentation in response to a request from the verifier. 
\end{itemize}

Presentation Exchange 2.0 is a specification for interaction between a verifier and an identity holder. Similar to the Aries Present Proof Protocol 2.0 is a transport agnostic specification. In addition, it is envelope agnostic, meaning it can be combined with different envelopes such as DIDComm v1, v2, or SIOP DID Profile 2.0. It can also be combined with varying types of transport like CHAPI or VC over HTTP. Moreover, it is credential format agnostic so that verifiable presentations and OIDC ID tokens\cite{presentation-exchange} can be exchanged with this specification. 

Another specification for interaction between an issuer and an identity holder is the Credential Manifest. It is a standard data format that describes the required information from the identity holder to issue a verifiable credential. Instead of specifying the states and message types seen in the Aries Issue Credential Protocol 2.0, the Credential Manifest specifies the JSON file that is passed to the identity holder that requests necessary information\cite{cred-manifest}. However, it is possible to use these JSON files as attachments with the Aries Issue Credential Protocol 2.0 and 3.0. 

On the other hand, Verifiable Presentation Request Specification v0.1 takes a different approach and specifies the query within a client application after a presentation exchange is requested. It is a JSON-based query whose results are wrapped in a VP. Verifiable Presentation Request can be used with Credential Handler API, Mobile Applications, and P2P interactions. Currently, it supports several query types such as Query By Example, DID Authentication Request, and Authorization Capability Request\cite{vp-request}.

Credential Handler API (CHAPI) is a browser application programming interface for wallet-less solutions that lets web applications securely use get and store functions for credentials without having an initial wallet infrastructure. CHAPI provides a secure and trusted user interface for managing credentials. It also gives the users the ability to choose service providers for their wallets\cite{chapi}.

Last but not least, the OpenID Connect Credential Provider is a specification that describes the workflows, interactions, and context of data exchanged between an issuer and an identity holder during the credential issuance process using the OIDC standard. It extends the existing specification OIDC protocol for an OpenID provider to issue credentials to the identity holders using JSON Web Signatures and JSON Web Encryption. The OpenID provider takes the role of a credential issuer, and the credential holder acts as a relying party. The end-user authenticates the credential issuer to issue credentials to the credential holder. This specification does not cover the interaction between the credential verifier and the credential holder\cite{oidc-cred-provider}.

\subsubsection{Different Credential Binding Mechanisms}

Depending on the credential format and proof, different approaches are taken. The first type of credential binding is based on binding by a DID: When a verifiable credential is issued to the identity holder, a DID is written in the credential subject section of the verifiable credential. This DID can also create secure communication using the DIDComm envelope. Since this DID is in control of the identity holder, it can also sign transactions. When a verifiable presentation is shown to a verifier, it is signed by the same public key that controls the DID written in the credential subject section of the verifiable credential\cite{vc-data-model}. This way, it is proven that the identity holder controls the public/private key pair used during the issuance process.

The second type of credential binding involves a secret known only to the identity holder. With ZKP, the holder sends a blinded version of this secret to the issuer. The issuer then signs it along with the other attributes of the credential\footnote{see Listing \ref{listing:jsoncl}: An Anonymous Credential with JSON format and CL Signature}. Since the issuer only sees the secret as a blind attribute, it does not know the exact value. Furthermore, due to the ZKP and selective disclosure, when the VP is presented, the identity holders prove that they are in possession of this secret without showing what it is. Since the same secret was being used as a blind attribute, the holders can prove the verifiable credential is bound to them without revealing a persistent identifier. This secret is called a link secret, also known as a master secret\cite{khovratovich2017sovrin}.

The last type of credential binding is used when using the SIOP ID token. In the case of federated identity solutions and an OpenID Provider, the ID token is generated by the OpenID Provider so that the origins are verified. There is no OpenID Provider with SSI, and identity holders can pass ID tokens themselves. When using non-self-attested credentials, proof of ownership and origins are required but cannot be proven by the ID token alone. OIDC4VP specifies that a VP token is passed along with the ID token in those cases. The VP token contains the verifiable presentation, underlying proofs, and credential binding methods. It means the credential binding is proven either by using a DID or a link secret\cite{oidc4vp}. 

Credential binding is a by-product of credential proof and is an indirect interoperability consideration: As long as interacting agents support identical credential proofs, the credential binding will follow.

\begin{figure*}[t]
    \centering
    \includegraphics[width=\textwidth]{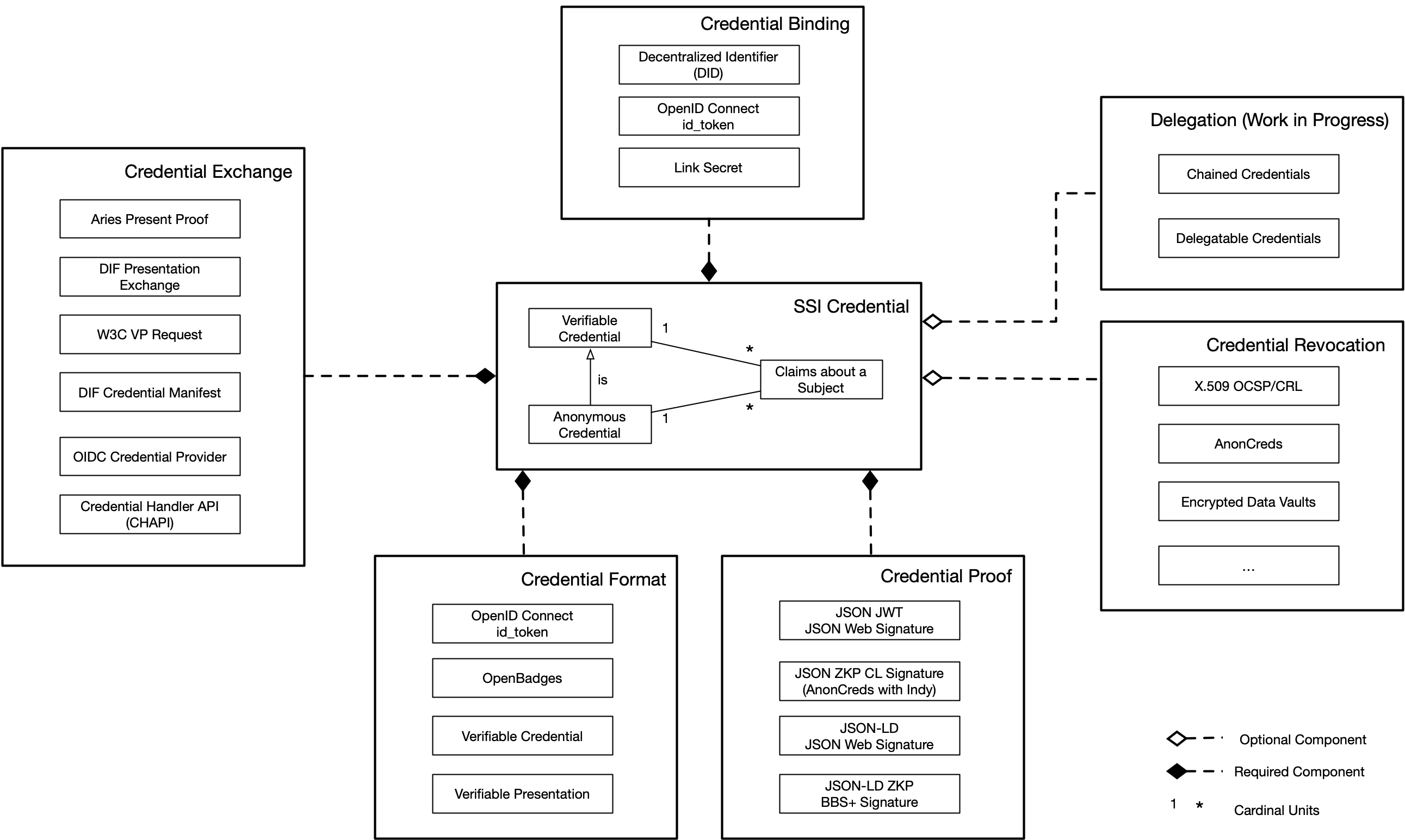}
    \caption{Credential Layer Composition}
    \label{fig:CredTax}
\end{figure*}

Figure \ref{fig:CredTax} depicts the components and the variations of those components that build up a self-sovereign identity. 

\subsection{Application Layer}
Interoperability considerations on the application layer tend to be less related to technical and syntactical interoperability since the underlying protocols and standards for exchanging self-sovereign identities are aligned in the first three layers of the reference model. Instead, interoperability considerations here are use case-specific and related to semantic interoperability. Semantic interoperability cannot be achieved in a general approach that has been seen in the previous layers, for example, by aligning the standards and protocols that agents follow. Instead, the alignment must happen on a credential level as it is the credential that enables a common understanding of the exchanged information between agents. This can be done in several ways:

\paragraph{JSON Credential Definitions}
One common method of creating semantic interoperability with credentials in JSON format is to create a structure that defines the exchanged credential and can be stored in a highly available location such as VDR. Then, when validation occurs, the VP refers to this credential definition so that the verifier knows which credential is being presented. The VP is based on this credential definition, and the verifier sees this as a context. Thus, semantic interoperability is achieved between the actors. The Hyperledger Indy technology stack follows this approach: issuers store schema definition and other verification material such as DIDs and public keys in the Indy ledger. Verifiers can create proof templates asking for one or more claims from that definition. 

\paragraph{JSON-LD Context}
The second method offers more flexibility by creating and issuing a VC that has a context maintained in an external domain such as schema.org\footnote{Welcome to schema.org!: \url{https://schema.org/}}. This context contains a vocabulary for the object. For example, a person can have multiple attributes, which can be mapped as structured data. At schema.org, a person has already been mapped. Using this mapping information to derive verifiable credentials is possible by using specific attributes defined at the Person type and linking the Person type from schema.org to the verifiable credential as a Linked Data context. For instance, if the Person type is linked to the VC via the use of \texttt{@context}, and when a claim like \texttt{birthPlace} is existent in this credential, it unequivocally means the place where the person was born \cite{schemaorgperson}. This is because "the place where the person was born" is the only description of \texttt{birthPlace} in that context. Linking structured data like this example is an efficient method of creating semantic disambiguity and shared understanding among the actors. 

Although not an interoperability problem between SSI technology stacks, with both methods, it is vital to keep considerations related to the vertical in mind to achieve adoption of the SSI paradigm by those verticals.

\subsection{Cross-Layer Considerations}
Interoperability considerations of the cross-layer components vary from technical to organizational interoperability, partially because these components are spread across all reference model layers. These interoperability considerations can be put into two categories. In the first category, just as in the reference model, the components must be aligned to achieve technical to organizational interoperability between SSI technology stacks. In the second category, cross-layer components are discussed to achieve interoperability with existing technology stacks, especially their identity access management systems, in verticals such as eHealth. This type of consideration is called interoperability with existing IDMS.

\subsubsection{Different Authentication and Authorization Protocols}
Coming from the federated identity paradigms, different authentication and authorization protocols are not interoperable with each other. When implementing use cases for verticals such as eHealth or education, it might be required to create transitional solutions with compatibility with those protocols to achieve interoperability with existing IDMS\cite{Yild2109:Connecting}. The predominant authentication and authorization protocols are discussed below. 

The W3C defines WebAuthn/FIDO2, a cryptographic protocol between a web application and a user\cite{webauth}. The corresponding FIDO2 API enables a web application the creation and use public key-based credentials, to strongly authenticate a user. FIDO2 enables the authentication of users with signed credentials and could be used as a replacement for username and password authentication. However, FIDO2 does not provide desktop, web, and mobile applications access to an API like a resource owner. 

For access to APIs between desktop, web, and mobile applications (relying parties) and resource owners, the combination of OIDC and OAuth 2.0 offers an alternative. OAuth 2.0 is a framework designed to support the development of authentication and authorization protocols. It provides a variety of standardized message flows based on JSON and HTTP\cite{sakimura2014openid} \cite{7956534}. OAuth 2.0 is widely adopted in verticals such as education. 

OIDC is an authentication protocol and identity layer on top of the OAuth 2.0 protocol widely adopted in verticals such as banking. OIDC allows obtaining basic profile information about relying parties in an interoperable and REST-like manner. Furthermore, the resource owner can verify the identity of relying parties based on the authentication performed by an authorization server. Technically, OIDC extends OAuth 2.0 and defines a method for authentication using a new type of token, the ID token.

A further extension of OAuth 2.0 is User-Managed Access (UMA) 2.0, a widely adopted protocol in eHealth, whose specification is published by the Kantara initiative \cite{uma}. UMA is an OAuth-based protocol designed to give resource owners a unified control point for authorizing who and what can get access to their digital data, content, and services, no matter where all those things live. Using UMA, the resource owner can configure an authorization server with authorization grant rules (policy conditions) at will, rather than authorizing access token issuance synchronously after authenticating. UMA thus needs a way to handle identity claims about the requesting party, and for this purpose, it integrates the option of using OIDC, which, like UMA, is based on OAuth 2.0. The UMA authorization server can act as an OIDC relying party to achieve this\cite{maler2017user}.

A different approach to OAuth 2.0 is the Grant Negotiation and Authorization Protocol (GNAP), which is currently being developed by an IETF working group \footnote{GNAP Working Group: https://datatracker.ietf.org/wg/gnap/about/}. While OAuth 2.0 uses client secrets and bearer tokens, GNAP seeks to move beyond that at the base level, using various security technologies. In GNAP, all communication between a relying party and the authorization server is bound to a key held by the relying party \footnote{XYZ: A set of implementations of GNAP: \url{https://oauth.xyz/keys/}}. GNAP uses the same cryptographic mechanisms to authenticate the relying party to the authorization server and bind the access token to the resource server and the authorization server. 




The last authorization method to be mentioned is based on authorization capabilities for Linked Data, published by the W3C credentials community group \footnote{Authorization Capabilities for Linked Data v0.3: \url{https://W3C-ccg.github.io/zcap-ld/}}. Authorization capabilities for Linked Data (ZCAP-LD for short) provide a secure way for Linked Data systems to grant and express authority utilizing the object-capability model. Koepe \cite{DBLP:journals/corr/abs-1907-07154} describes the object-capability model as a security measure that encodes access rights in individual objects to restrict their interactions with other objects. Capabilities are represented as Linked Data objects signed with data integrity proofs. In contrast to the previously described authorization methods, ZCAP-LD focuses not on the identity but on whether the user has the key to access the requested resource.


\subsubsection{Different Zero-Knowledge Proofs}
The SSI community is striving for unified selective disclosure of claims through ZKP. As some of them are explained in proof types, several ZKPs can enable selective disclosure. In the following, we discuss different ZKP approaches and their interoperability considerations. 

The zk-SNARKS and zk-STARKS belong to the group of general-purpose ZKP systems. The latter work with polynomials and hash functions and are thus currently considered post-quantum secure. The first and most zk-SNARKS constructions are based on pairings-based cryptography. They suffer from relatively poor performance since many pairings were needed. In addition, some of these constructions were and are criticized concerning security considerations since, e.g., a trusted setup phase is required to keep the toxic waste created during the generation of the Common Reference String (CRS) secret. Toxic waste includes private parameters that can be used to forge proofs.  However, recent developments of zk-SNARKS also use entirely different cryptographic primitives, such as lattice cryptography \cite{Gennaro_2018} or eliminate the need for a trusted setup in their construction \cite{Wahby_2018,cryptoeprint:2019:550}. Thus, there are already severe technical differences within the zk-SNARK constructions or between zk-SNARKS and zk-STARKS, which lead to non-interoperable systems on the technical level in the first place.

The signature-based ZKP procedures such as the CL Signatures \cite{Camenisch_2003} and BBS+ Signatures \cite{BBS-sigs-2001,Au_2006} also rely on different cryptographic primitives. While the CL Signatures use RSA mechanics and build on the Strong RSA Assumption \cite{Boneh2004}, BBS+ uses pairing-based cryptography and a bilinear mapping.  BBS+ Signatures, therefore, require a pairing-friendly elliptic curve. From a security point of view, BBS+ is secure in the standard model under the q-strong Diffie-Hellman (q-SDH) assumption \cite{Au_2006}.  

In principle, all these different classes of ZKP approaches and constructions are not interoperable on a technical level. However, assuming that freely available libraries are available for the use of a particular ZKP construction, technical interoperability can be achieved. 


Interoperability also means standardization by standards developing organizations (SDOs) such as IETF. AnonCreds and BBS+ Signatures are not standardized yet. Only peer-reviewed scientific publications and early drafts of standards exist at the moment. An informal BBS+ specification has been published by Mattr Global on GitHub \cite{bbs+spec}. Pairing-friendly elliptic curves are also just starting to be standardized. ISO is currently working on finalizing its draft ISO/IEC DIS 15946-5:2021, which also takes pairing-friendly curves into account. Also, the IETF currently has a draft in version 10 under its editing \cite{IETF-pairing-curves}. 



\subsubsection{Different Compliance Requirements}
Compliance is crucial to achieving organizational interoperability, especially legal interoperability. In the reference model, we discussed only GDPR as a local regulation to comply with. However, besides many other regional regulations, there are also legal trust frameworks whose requirements must be fulfilled for VCs to achieve a high level of assurance, which might be required for the use cases that involve electronic identifications. These legal trust frameworks vary across regions as well. Therefore, to achieve organizational interoperability between different technology stacks, regional regulations and, if required, trust frameworks must be supported by the interacting agents. 

\subsubsection{Different Data Formats}
Except for some edge cases such as YAML and JSON, data formats are incompatible, meaning the receiving agent supporting different data formats than the sender agent cannot interpret the incoming information. Therefore, the support of divergent data formats can hinder syntactical interoperability. However, this hindrance can be avoided by either aligning the supported data formats between agents or implementing data format converters. For instance, programming languages such as Java support the conversion of XML to JSON natively. If an application receives a non-native data format, it can be converted to the natively supported data format in the application logic. 

\subsubsection{Different Crypto Primitives}
There are numerous crypto primitives that technology stacks can include to perform SSI-related operations. From an interoperability perspective, crypto primitives play a substantial role as the same crypto primitives must be supported in the crypto libraries of the communicating agents. Otherwise, credentials cannot be issued, presented proofs cannot be interpreted, or information cannot be exchanged securely. 

To achieve technical interoperability on the crypto primitive level while the agents rely on divergent crypto libraries, it is crucial to take the considerations of implementation guidelines mentioned in the signature suites or community drafts, i.e., the signature suites for BBS+\footnote{BBS+ Signatures 2020: \url{https://W3C-ccg.github.io/ldp-bbs2020/}} or Edwards-Curves\footnote{Ed25519 Signature 2018: \url{https://W3C-ccg.github.io/lds-ed25519-2018/}}. In the case of BBS+ Signatures, a BBS+ signature only counts as a data integrity proof capable signature, according to the community draft specification of W3C, when it is paired with the BLS12-381 curve, which is one of the subsets of elliptic curves known to be pairing friendly\cite{yonezawa2019pairing}\cite{pairing-friendly-curves}.



\section{Conclusion} \label{conclude}
The latest digital identity paradigm SSI is gaining much traction regarding the number of implementations and expectations. For self-sovereign identities to be truly self-sovereign, they must be widely accepted and, therefore, interoperable with most service providers. However, the current landscape proves that the technology stacks are not interoperable with each other. Without interoperability, wide adoption with identity subjects in focus is not fathomable. In this tutorial, we first created a definition based on existing models to have a common understanding of interoperability between entities. Moreover, we presented a reference model for an understanding of differences between technology stacks which then can be used for aligning those stacks in components required to be compatible to achieve the interoperability goals. 

The SSI landscape is constantly and rapidly changing. Some protocols and standards are becoming obsolete while others are recently being created. Therefore, the protocols and standards mentioned in components of the reference model in this tutorial are not to be seen as final but more as a guideline to understand the differences between technology stacks. As long as no one technology stack fulfills every business requirement, multiple technology stacks will exist and might be compatible with each other. Therefore, when an alignment is needed, and SSI technology stack creators understand their differences related to interoperability with the reference model, they should do due diligence about the current protocols and standards related to the component that must be aligned.

\section*{Acknowledgment}

Although we do not know half of the persons mentioned below on a personal level, we would like to thank Daniel Hardman, Stephen Curran, Kaliya Young, Marcus Sabadello, Daniel Buchner, Dave Longley, Tobias Looker, Orie Steele, Manu Sporny, along with many other writers and countless editors who contributed so much to the SSI community. Furthermore, we would like to thank Mirko Mollik, Artur Philipp, Sebastian Schmittner, Ralph Tröger, and Sebastian Zickau for their contributions to the IDunion working group for the SSI component investigation.

\bibliography{bibliography.bib}

\begin{thebibliography}{100}
\bibitem{righttobeforgotten}
``{}backups and the right to be forgotten in the gdpr: An uneasy
  relationship.''
\end{thebibliography}
\bibliographystyle{IEEEtran}

\end{document}